\documentclass[traditabstract]{aa}
\usepackage{rotating}
\usepackage{graphicx}
\usepackage{txfonts}
\usepackage{natbib}
\usepackage{epsfig}

\usepackage{amstext}

\begin{document}

\title{Apsidal motion in the massive binary HD~152218}
\author{G.\ Rauw\inst{1} \and S.\ Rosu\inst{1} \and A.\ Noels\inst{1} \and L.\ Mahy\inst{1}\fnmsep\thanks{Postdoctoral Researcher FRS-FNRS (Belgium)} \and J.H.M.M.\ Schmitt\inst{2} \and M.\ Godart\inst{1} \and M.-A.\ Dupret\inst{1} \and E.\ Gosset\inst{1}\fnmsep\thanks{Senior Research Associate FRS-FNRS (Belgium)}}
\offprints{G.\ Rauw}
\mail{rauw@astro.ulg.ac.be}
\institute{Space sciences, Technologies and Astrophysics Research (STAR) Institute, Universit\'e de Li\`ege, All\'ee du 6 Ao\^ut, 19c, B\^at B5c, 4000 Li\`ege, Belgium \and Hamburger Sternwarte, Universit\"at Hamburg, Gojenbergsweg 112, 21029 Hamburg, Germany}
\date{}
\abstract{Massive binary systems are important laboratories in
which to probe the properties of massive stars and stellar physics in general. In this context, we analysed optical spectroscopy and photometry of the eccentric short-period early-type binary HD~152218 in the young open cluster NGC~6231. We reconstructed the spectra of the individual stars using a separating code. The individual spectra were then compared with synthetic spectra obtained with the CMFGEN model atmosphere code. We furthermore analysed the light curve of the binary and used it to constrain the orbital inclination and to derive absolute masses of $(19.8 \pm 1.5)$ and $(15.0 \pm 1.1)$\,M$_{\odot}$. Combining radial velocity measurements from over 60 years, we show that the system displays apsidal motion at a rate of $(2.04^{+.23}_{-.24})^{\circ}$\,yr$^{-1}$. Solving the Clairaut-Radau equation, we used stellar evolution models, obtained with the CLES code, to compute the internal structure constants and to evaluate the theoretically predicted rate of apsidal motion as a function of stellar age and primary mass. In this way, we determine an age of $5.8 \pm 0.6$\,Myr for HD~152218, which is towards the higher end of, but compatible with, the range of ages of the massive star population of NGC~6231 as determined from isochrone fitting.} 
\authorrunning{Rauw et al.}
\titlerunning{Apsidal motion of HD~152218}
\maketitle


\section{Introduction}
In recent years, it has been found that the majority of the massive stars belong to binary or higher multiplicity systems \citep[][and references therein]{SF}. This situation has considerable implications for our understanding of the evolution of massive stars, but at the same time, it also offers enormous possibilities to observationally constrain the properties of these stars. Particularly interesting are unevolved double-line spectroscopic eclipsing binary systems (so-called SB2Es). The joint analysis of the photometric eclipses and radial velocities inferred from spectroscopy allows us to determine absolute masses and radii that are essentially model independent. Moreover, for unevolved systems, the properties of these stars should be good proxies of those of single massive stars. 

Of special interest are eccentric SB2Es that show significant apsidal motion \citep[e.g.][]{Bulut,Schmitt}. The latter arises from the fact that the gravitational field of stars in a close binary system can no longer be approximated as the gravitational field of a point-like mass. As a result, the orbits of the stars can only to first order be described as closed ellipses. A better description is achieved when it is considered that the argument of periastron $\omega$ undergoes a secular perturbation $\dot{\omega}$ \citep[e.g.][and references therein]{Schmitt}. The rate of apsidal motion is directly related to the internal structure of the components of the binary \citep[e.g.][]{Shakura,CG92,CG10}, thus allowing us to obtain information on the internal mass-distribution of the stars. Measuring the rate of apsidal motion can also provide an estimate of the masses of the components of non-eclipsing eccentric close binaries \citep{Benvenuto,Ferrero}, although this method is strongly model dependent. To date, there are only very few massive stars for which detailed studies of apsidal motion have been performed. Only four systems among the 128 eccentric eclipsing binaries listed by \citet{Bulut} host at least one star with a mass of more than 20\,M$_{\odot}$. In this paper, we discuss a fifth system: HD~152218.

HD~152218 is a member of the young open cluster NGC~6231 in the Sco\,OB1 association. With a binary fraction of at least 0.63, the O-star population of this cluster appears to be very rich in binary systems, especially short-period systems \citep{NGC6231}. HD~152218 is a binary system consisting of an O9\,IV primary and an O9.7\,V secondary \citep{Sana}. The system has a rather high eccentricity ($e \simeq 0.28)$ given its orbital period of about 5.604\,days \citep{Hill,Stickland,Sana}. \citet{Otero} reported HD~152218 to display shallow eclipses. They classified the system as an Algol-type eclipsing binary, but did not perform a detailed analysis of its light curve. Comparing their radial-velocity curve obtained from data collected between 1997 and 2004 with previously published solutions from the literature \citep{Struve,Hill,Stickland}, \citet{Sana} noted the probable presence of a significant apsidal motion with $\dot{\omega}$ between 1.4 and 3.3$^{\circ}$\,yr$^{-1}$, depending on the data sets that were considered. {\it XMM-Newton} data furthermore revealed a modulation in the X-ray flux of HD~152218 with orbital phase that was interpreted as a result of a wind-wind interaction \citep{Sana,GRYN}. 

In the present paper, we re-address the issue of apsidal motion in the HD~152218 binary system. In Sect.\,\ref{observations} we introduce the observational data. The radial velocities and individual spectra of the binary components are analysed in Sect.\,\ref{spectroscopy}, notably to establish the rate of apsidal motion. Section\,\ref{photometry} presents our analysis of the photometric light curve, whilst Sect.\,\ref{theory} compares the observed rate of apsidal motion to predictions from theoretical models. Finally, in Sects.\,\ref{discussion} and \ref{conclusions} we discuss our results and present our conclusions. 

\section{Observational data \label{observations}}
\subsection{Spectroscopic data \label{specobs}}
Our spectroscopic analysis is based on 28 high-resolution \'echelle spectra obtained with the FEROS spectrograph at ESO between April 1999 and May 2004. Details on these observations and the data reduction are given in \citet{Sana}. To derive the radial velocities (RVs) and to reconstruct the individual spectra of the primary and secondary components of HD~152218, we applied our spectral disentangling code, which is based on the method outlined by \citet{GL}. In this procedure, the individual spectra are reconstructed in an iterative way by averaging the observed spectra shifted into the frame of reference of one binary component after subtracting the current best approximation of the companion spectrum shifted to its current estimated radial velocity. Improved estimates of the RVs of the stars are obtained by cross-correlating the residual spectra, obtained after subtracting the companion spectrum, with a mask that is equal to zero in the continuum and equal to one at the rest wavelengths of spectral lines. The revised RVs are then computed from the centroid of the cross-correlation peak. This approach is especially powerful near conjunction phases, where conventional Gaussian fits usually fail to properly deblend the lines. As an initial approximation of the RVs, we used the values derived by \citet{Sana}. Our newly derived RVs are used in the coming sections to constrain the physical parameters of the system.
\subsection{Photometric data \label{photobs}}
 We have extracted $V$-band photometry of HD~152218 taken from the All Sky Automated Survey \citep[ASAS-3,][]{ASAS} carried out at the Las Campanas Observatory in Chile. The ASAS-3 system consisted of two wide-field ($8.8^{\circ} \times 8.8^{\circ}$) telescopes, each equipped with a 200/2.8 Minolta telephoto lens and a $2048 \times 2048$ pixels AP-10 CCD camera. These instruments were complemented by a narrow-field ($2.2^{\circ} \times 2.2^{\circ}$) 25\,cm Cassegrain telescope equipped with the same type of CCD camera.

The ASAS-3 photometric catalogue provides magnitude measurements performed with five different apertures varying in diameter from 2 to 6 pixels. Substantial differences between large- and small-aperture magnitudes indicate either contamination by close neighbours, in which case large apertures should be avoided, or saturation. The ASAS-3 data are not uniform in terms of exposure time and thus in terms of saturation limit, as the exposure time was changed in the course of the project from 180\,s (saturation near $V \sim 7.5$) to 60\,s (saturation around $V \sim 6$). The saturation limit also depends on the quality of the focus and the zenith distance. The ASAS-3 catalogue lists a grade for each epoch at which the star was observed. However, this grade is calculated for the frame as a whole, not for a specific object inside the field of view. Hence, a frame can have grade A, but still contain stars that saturated.

We have filtered the ASAS-3 data of HD~152218, keeping only data with a quality flag A and a mean error on the photometry of less than 0.034\,mag. Since HD~152218 has a $V$ magnitude close to the saturation limit of the ASAS instruments, we requested in addition that the dispersion between the various aperture photometric measurements be less than the mean error on the photometry and that the $V$ -band magnitudes should be in the range between 7.55 and 7.80. This procedure resulted in a light curve with a total of 126 photometric data points collected between HJD~2\,451\,963 and HJD~2\,454\,707.\\ 

\section{Spectral analysis \label{spectroscopy}}
\subsection{RV analysis}
In close eccentric binaries, the non-spherical shape of the stars induces a secular variation of the longitude of periastron ($\omega$) known as apsidal motion. This in turn leads to slow variations in the shape of the RV curve. Before we analyse the RVs to derive the orbital parameters and constrain the apsidal motion of HD~152218, we briefly consider the observational consequences of apsidal motion. 

As a result of the apsidal motion, the notion of the orbital period becomes ambiguous and different definitions of the period can be adopted \citep[see][for a discussion of this problem]{Schmitt}. A practical consequence is that the Fourier period search carried out on data taken over several years, such as performed by \citet{Sana}, is biased. Since the apses move in the same sense as the orbital motion, we expect the Fourier technique to underestimate the value of the period between consecutive periastron passages. Likewise, apsidal motion will also affect the value of the eccentricity determined from a fit of data spanning several years. The apsidal motion smears out the peaks of the RV curve, thus leading to an apparently lower value of $e$. The orbital period, eccentricity, longitude of periastron, and time of periastron passage derived from a set of RVs collected over several years are therefore most likely biased by apsidal motion, whilst the amplitudes $K_1$ and $K_2$ should be free of such biases.

To establish the amplitudes of the RV curves, we hence computed an orbital solution applying the Li\`ege Orbital Solution Package (LOSP) code \citep{SGR}, which is an improved version of the code originally proposed by \citet{WHS}, to the set of RV values obtained in Sect.\,\ref{specobs}. The resulting set of orbital parameters (see Col.\ 3 of Table\,\ref{bestfit}) was used as a first guess of the parameters to determine the rate of apsidal motion. 

To establish the orbital solution accounting for apsidal motion, we considered the radial velocities of the primary available in the literature. Our analysis includes the photographic measurements presented by \citet{Struve} with the heliocentric Julian dates computed in the same way as indicated in \cite{Sana}, the photographic measurements of \citet{Hill}, the {\it IUE} data of \citet{Stickland}, the CAT+CES data from \citet{Sana}, and our newly determined radial velocities from the FEROS spectra. To establish the RVs of the components of HD~152218, \citet{Stickland} first brought the {\it IUE} spectra to an almost absolute frame of reference using the interstellar lines. The RVs were obtained through cross-correlation with comparison stars. For phases where the lines of the two stars were not fully deblended, the widths of the individual cross-correlation functions were fixed to the values obtained at phases when the lines were resolved. See \citet{Stickland} for further details. We did not consider the RV data from \citet{GM01} because they were found to be subject to clerical errors \citep[see discussion in][]{Sana}. All RV data were assigned an error that was set to the mean error about the RV solution obtained for the corresponding epoch.

For each time of observation $t$, we adjusted these RV data with the following relation
\begin{equation}
\label{eq:RV}
RV_1(t) = \gamma_1 + K_1\,[\cos{(\phi(t)+\omega(t))} + e\,\cos{\omega(t)}]
,\end{equation}
where 
\begin{equation}
\label{eq:w}
\omega(t) = \omega_0 + \dot{\omega}\,(t - T_0)
.\end{equation}
$\gamma_1$ is the apparent systemic velocity of the primary, which was set to the value found for the best orbital solution for that epoch. $K_1$ is the amplitude of the primary RV curve, which we set to 160.3\,km\,s$^{-1}$ as this was the value consistently found for all of the epochs. $\phi(t)$ stands for the true anomaly computed by solving Kepler's equation with a specific value of the orbital period $P_{\rm orb}$, the eccentricity $e$, and the time of periastron passage $T_0$. $\omega$ is the argument of periastron, $\omega_0$ is the value of $\omega$ at time $T_0$, and $\dot{\omega}$ is the rate of secular apsidal motion. 

\begin{figure}[htb]
\begin{center}
\includegraphics*[width=0.47\textwidth,angle=0]{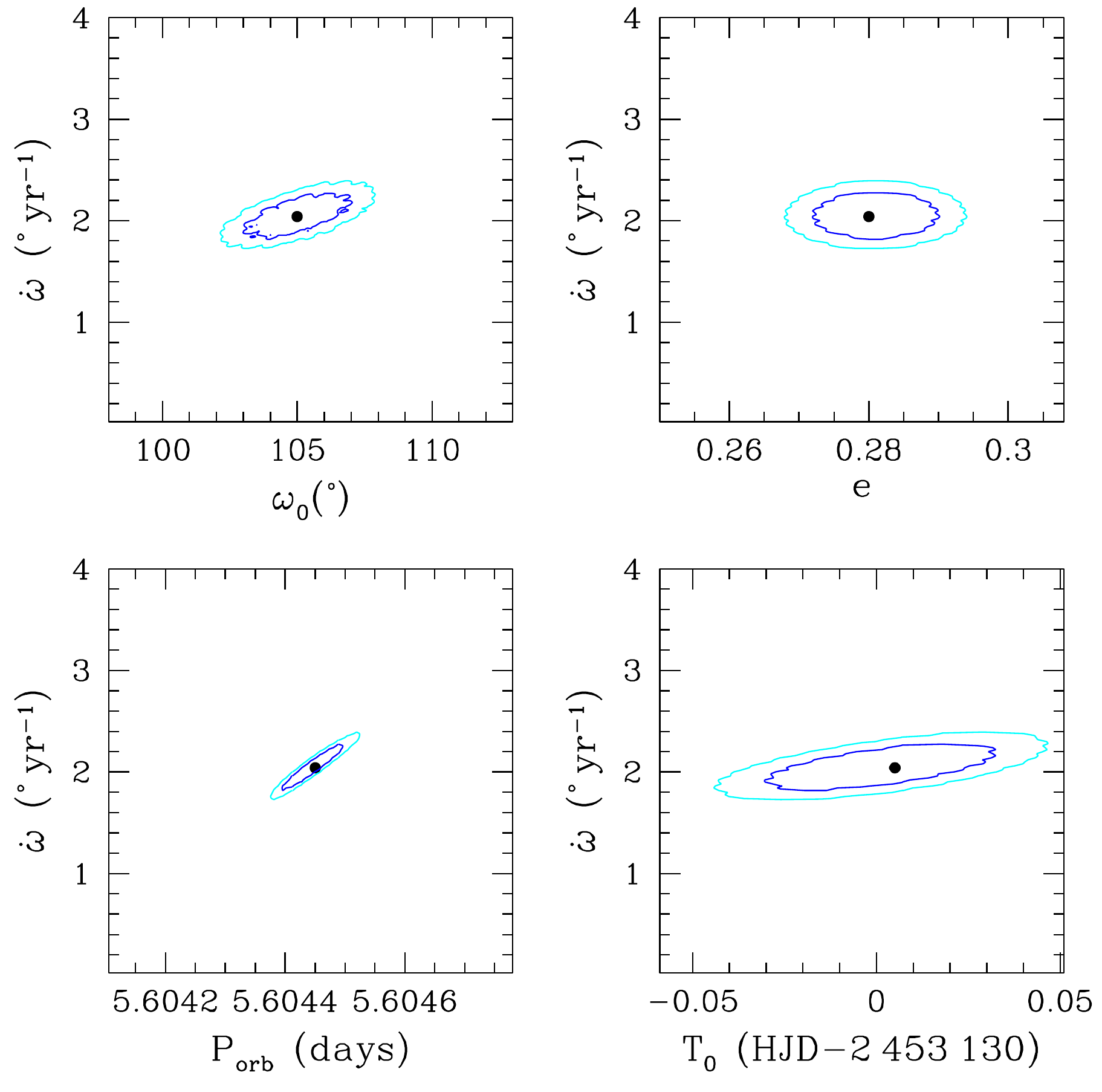}
\end{center}
\caption{Confidence contours for the best-fit parameters of the adjustment of the full set of 81 RV data of the primary star with Eqs.\ \ref{eq:RV} and \ref{eq:w}. The blue and cyan contours correspond to $1\,\sigma$ and 90\% confidence levels, respectively. The best-fit solution is indicated by the filled symbol.\label{contoursomega}}
\end{figure}

This therefore is a problem with five free parameters: $P_{\rm orb}$, $e$, $T_0$, $\omega_0$, and $\dot{\omega}$. We scanned the parameter space in a systematic way, starting near the orbital solution obtained from the FEROS data. Some examples of projections of the five-dimensional parameter space onto 2D planes are shown in Fig.\,\ref{contoursomega}.

In this way, we found the best-fit parameters and their $1\sigma$ errors, which are listed in Col.\ 4 of Table\,\ref{bestfit}.

\begin{figure}[htb]
\begin{center}
\includegraphics*[width=0.45\textwidth,angle=0]{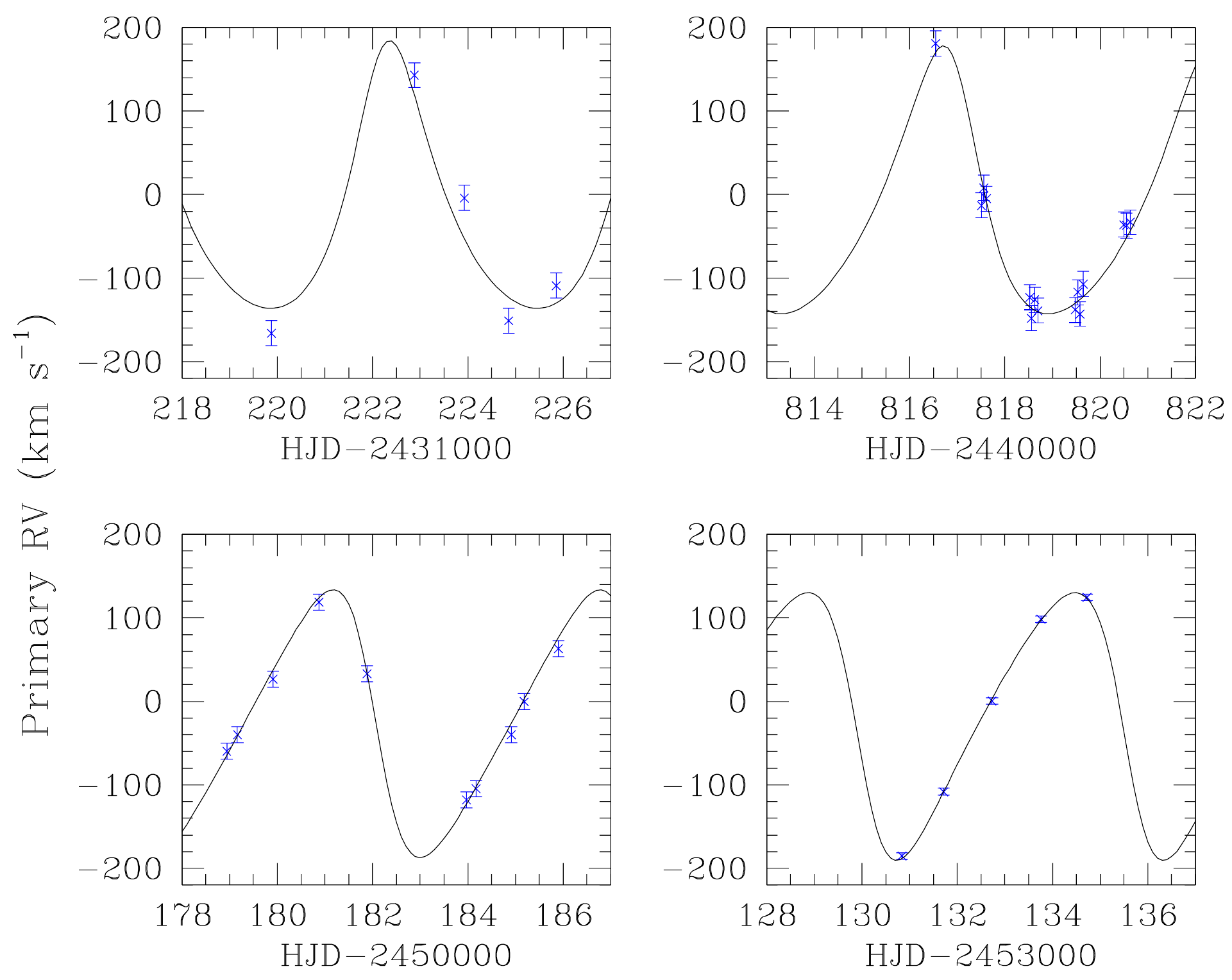}
\end{center}
\caption{Examples of the comparison between the RV data of the primary star and relations \ref{eq:RV} and \ref{eq:w} with the best-fit parameters given in Col.\ $[4]$ of Table\,\ref{bestfit}. The top panels correspond to data from \citet[][left]{Struve} and \citet[][right]{Hill}. The bottom panels show one epoch of {\it IUE} data from \citet[][left]{Stickland} and one epoch of FEROS data presented by \citet{Sana} and reanalysed here (right). \label{fitRV}}
\end{figure}

As an example, Fig.\,\ref{fitRV} illustrates the fit of the radial velocity data at four different epochs. The change in morphology of the RV curve over 60 years between the observations of \citet{Struve} and \citet{Sana} is evident.

Of course, the photographic data of \citet{Struve} and, to a lesser extent, of \citet{Hill} are of lower quality than the more recent {\it IUE} and optical echelle spectra. We therefore also applied the same technique to the 61 RV points consisting of the \citet{Stickland} data, the CES RVs from \citet{Sana} and our newly determined RVs for the FEROS data. The results are in good agreement with those of the full set of RV data (see Col.\ 5 of Table\,\ref{bestfit}), although, of course, with slightly larger error bars. This test demonstrates that our determination of the rate of apsidal motion is reliable. 

\begin{table*}[htb]
\caption{Best-fit orbital parameters of HD~152218.}
\begin{center}
\begin{tabular}{c c c c c c}
\hline\hline
Parameter & \citet{Stickland} & \citet{Sana} & \multicolumn{3}{c}{This work} \\
\cline{4-6}
          &                   &              & New FEROS RVs & All RV data & Echelle spectra \\
& $[1]$ & $[2]$ & $[3]$ & $[4]$ & $[5]$ \\
\hline
\vspace*{-3mm}\\
$P_{\rm orb}$\,(d) & $5.603979$ & $5.60391$ & $5.60400 \pm 0.00004$ & $5.60445^{+.00005}_{-.00005}$ & $5.60447^{+.00011}_{-.00014}$ \\
\vspace*{-3mm}\\
$e$              & $0.308 \pm 0.018$ & $0.259 \pm 0.006$ & $0.269 \pm 0.005$ & $0.280^{+.010}_{-.008}$ & $0.284^{+.009}_{-.010}$\\
\vspace*{-3mm}\\
$\omega$\,($^{\circ}$) & $80.6 \pm 2.1$  & $104.0 \pm 1.6$ & $103.4 \pm 0.9$ &       &  \\
\vspace*{-3mm}\\
$\dot{\omega}$($^{\circ}$\,yr$^{-1}$) &   & $1.4$ -- $3.2$ &  & $2.04^{+.23}_{-.24}$ & $1.98^{+.59}_{-.56}$\\
\vspace*{-3mm}\\
$\omega_0$ ($^{\circ}$) &  & & & $105.0^{+2.0}_{-2.0}$ & $105.0^{+2.4}_{-2.9}$ \\
\vspace*{-3mm}\\
$T_0$ (HJD$-$2\,400\,000) & $48\,865.024 \pm 0.028$ & $53\,197.229 \pm 0.025$ & $53\,129.986 \pm 0.015$ & $53\,130.005^{+.028}_{-.036}$ &$53\,130.009^{+.041}_{-.041}$\\
\vspace*{-3mm}\\
$q = m_1/m_2$      & $1.341 \pm 0.030$ & $1.319 \pm 0.014$ & $1.320 \pm 0.011$ & & \\
$K_1$\,(km\,s$^{-1}$) & $156.9 \pm 3.6$ & $162.2 \pm 1.2$ & $160.3 \pm 1.0$ & & \\
$K_2$\,(km\,s$^{-1}$) & $210.4 \pm 3.5$ & $213.9 \pm 1.5$& $211.5 \pm 1.3$ & & \\
$\gamma_1$\,(km\,s$^{-1}$) & $-26.6 \pm 1.9$ & $0.9 \pm 1.2$ & $-18.6 \pm 0.8$ & & \\
$\gamma_2$\,(km\,s$^{-1}$) & $= \gamma_1$ & $-0.6 \pm 1.4$ & $-19.3 \pm 0.9$ & & \\
$a_1\,\sin{i}$\,(R$_{\odot}$) & $16.53 \pm 0.39$ & $17.34 \pm 0.13$ & $17.1 \pm 0.1$ & & \\
$a_2\,\sin{i}$\,(R$_{\odot}$) & $22.17 \pm 0.39$ & $22.87 \pm 0.17$ & $22.6 \pm 0.1$ & & \\
$m_1\,\sin^3{i}$\,(M$_{\odot}$) & $14.22 \pm 0.64$ & $15.82 \pm 0.26$ & $15.2 \pm 0.2$ & & \\
$m_2\,\sin^3{i}$\,(M$_{\odot}$) & $10.60 \pm 0.53$ & $12.00 \pm 0.19$ & $11.5 \pm 0.2$ & & \\
\hline
\end{tabular}
\end{center}
\tablefoot{Columns 1 and 2 list the orbital solutions from the literature for comparison with our results. The last two columns yield the best-fit parameters accounting for apsidal motion for the full set of 81 RV measurements of the primary star and for the 61 \'echelle spectra only.}
\label{bestfit}
\end{table*} 
\begin{figure*}[h!tb]
\begin{center}
\includegraphics*[width=0.45\textwidth,angle=0]{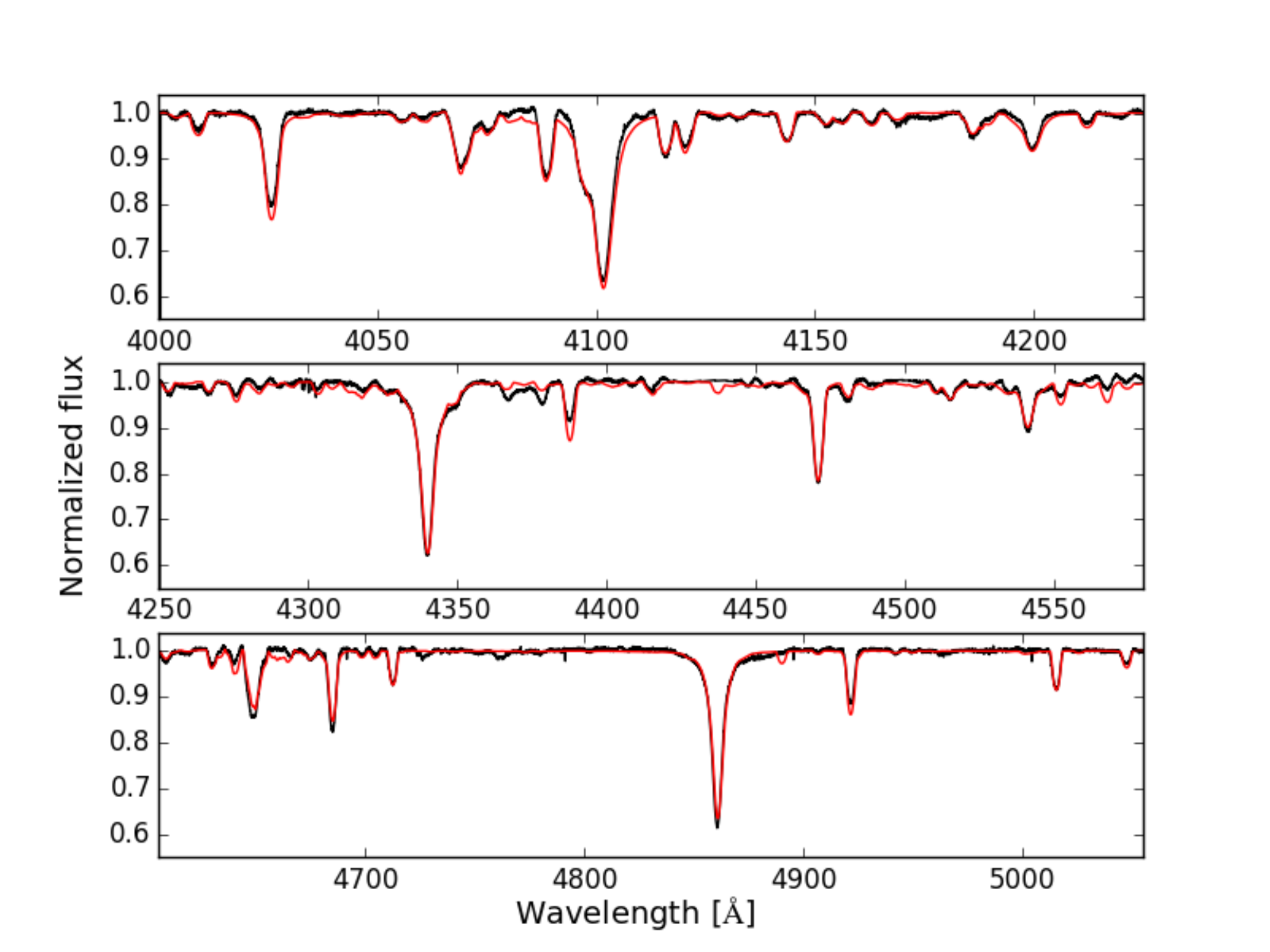}
\includegraphics*[width=0.45\textwidth,angle=0]{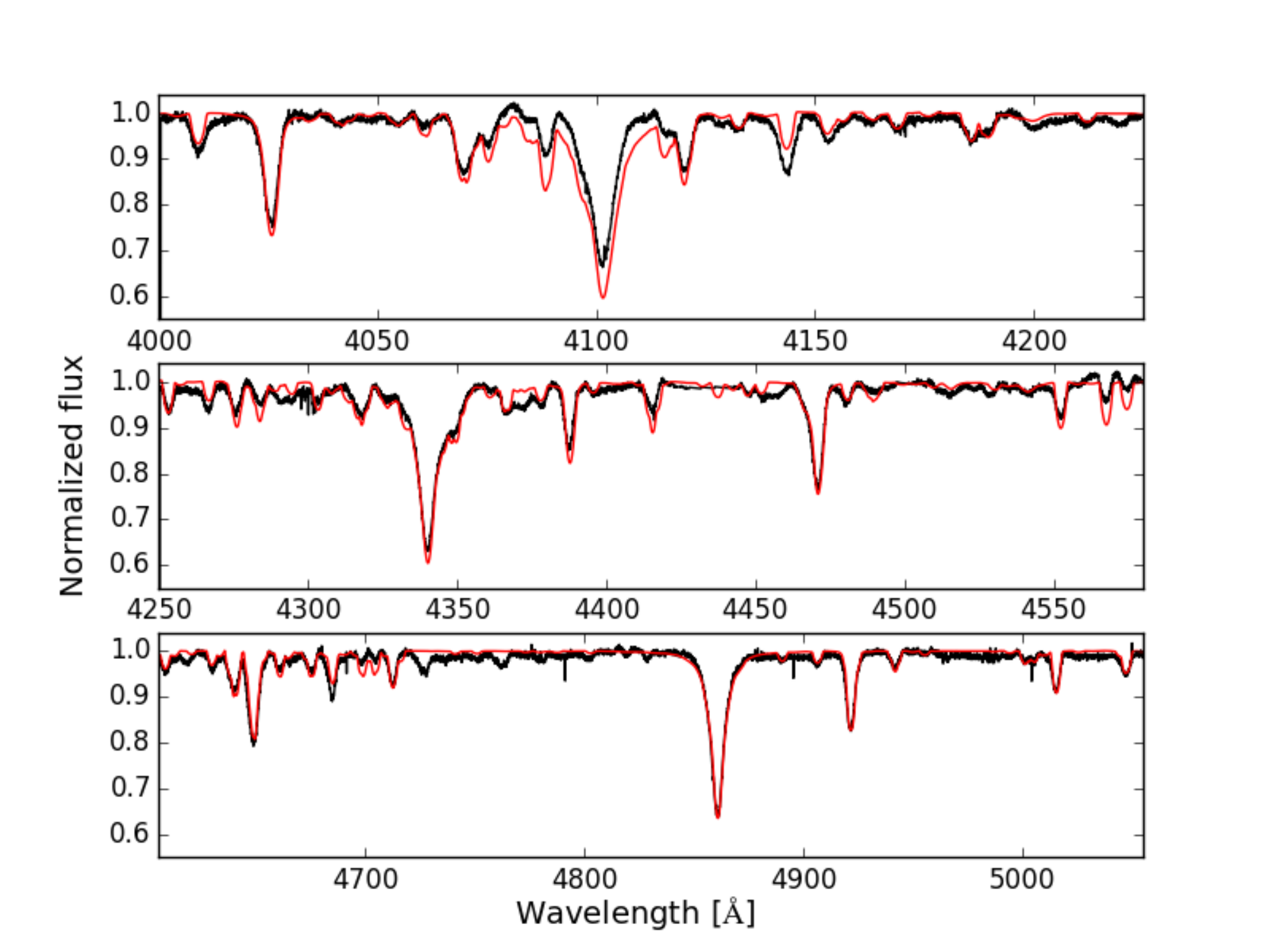}
\includegraphics*[width=0.45\textwidth,angle=0]{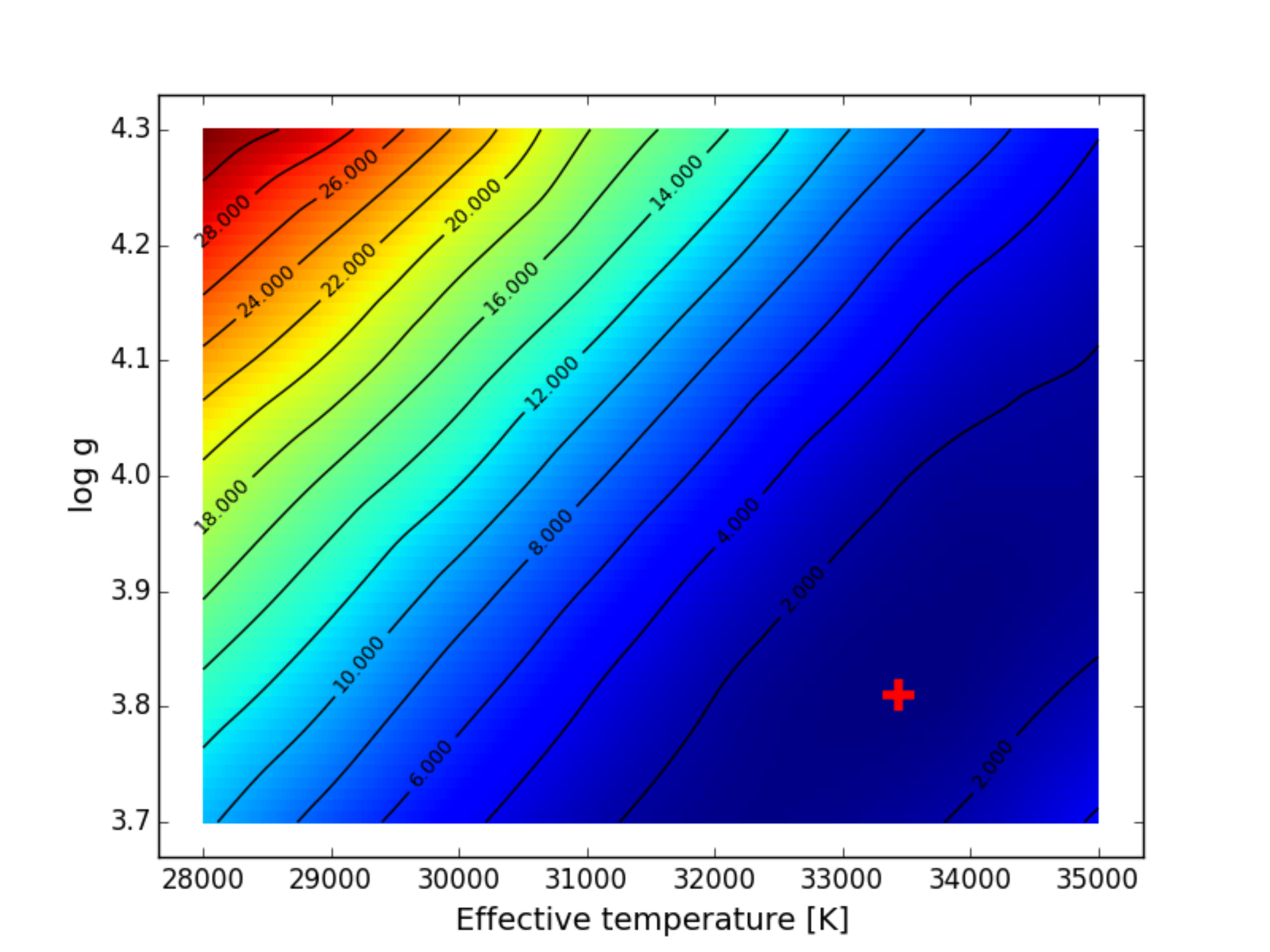}
\includegraphics*[width=0.45\textwidth,angle=0]{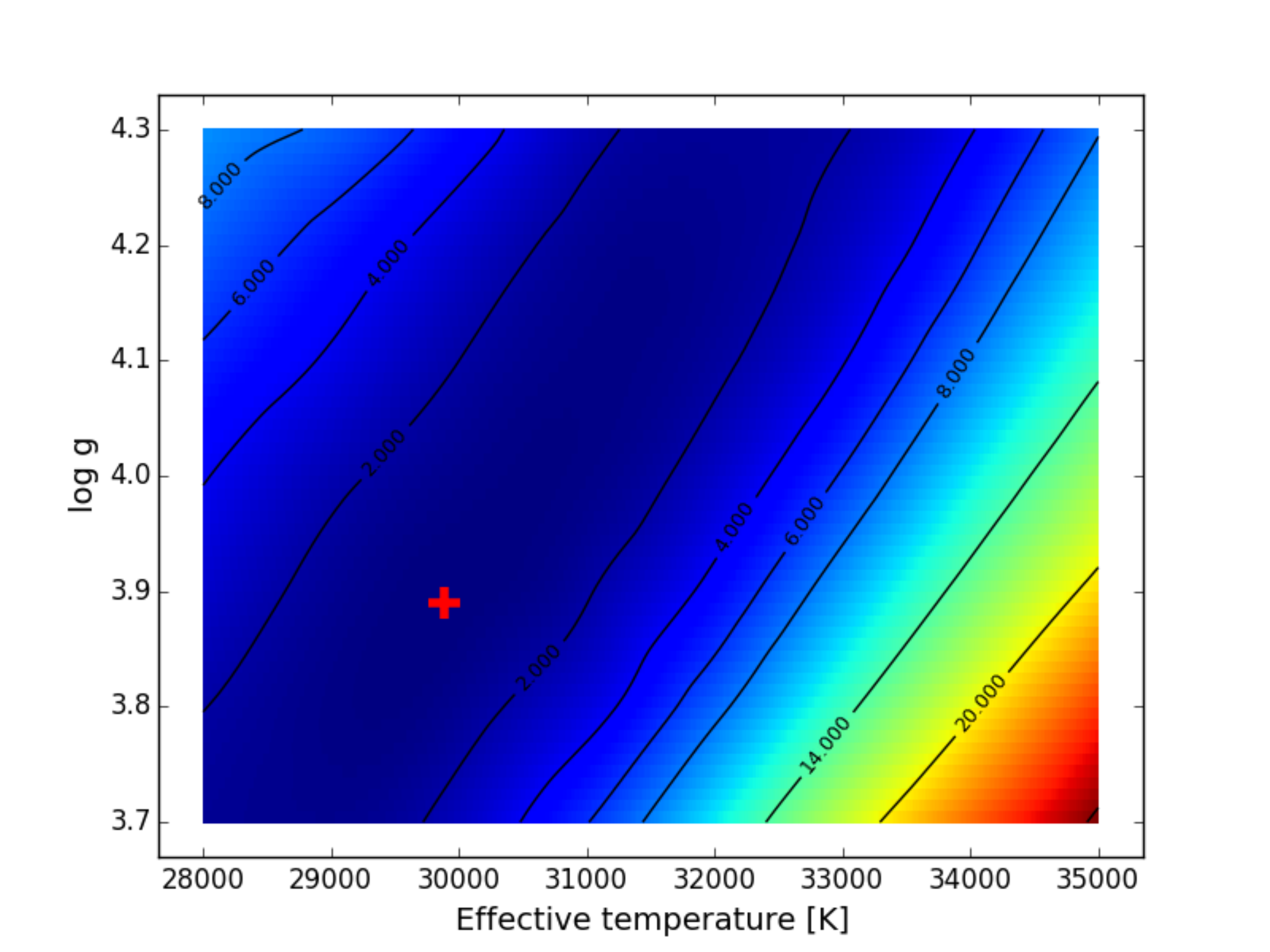}
\end{center}
\caption{Top row: comparison between the primary (left) and secondary (right) spectra obtained with the disentangling process (black lines) and the best-fit CMFGEN models (red lines). Bottom row: $\chi^2$ map of the CMFGEN fits showing the errors on our best-fit temperatures and gravities (shown by the red crosses) for the primary (left) and secondary (right).\label{specfit}}
\end{figure*}  

\begin{figure*}
\begin{center}
\includegraphics*[width=0.45\textwidth,angle=0]{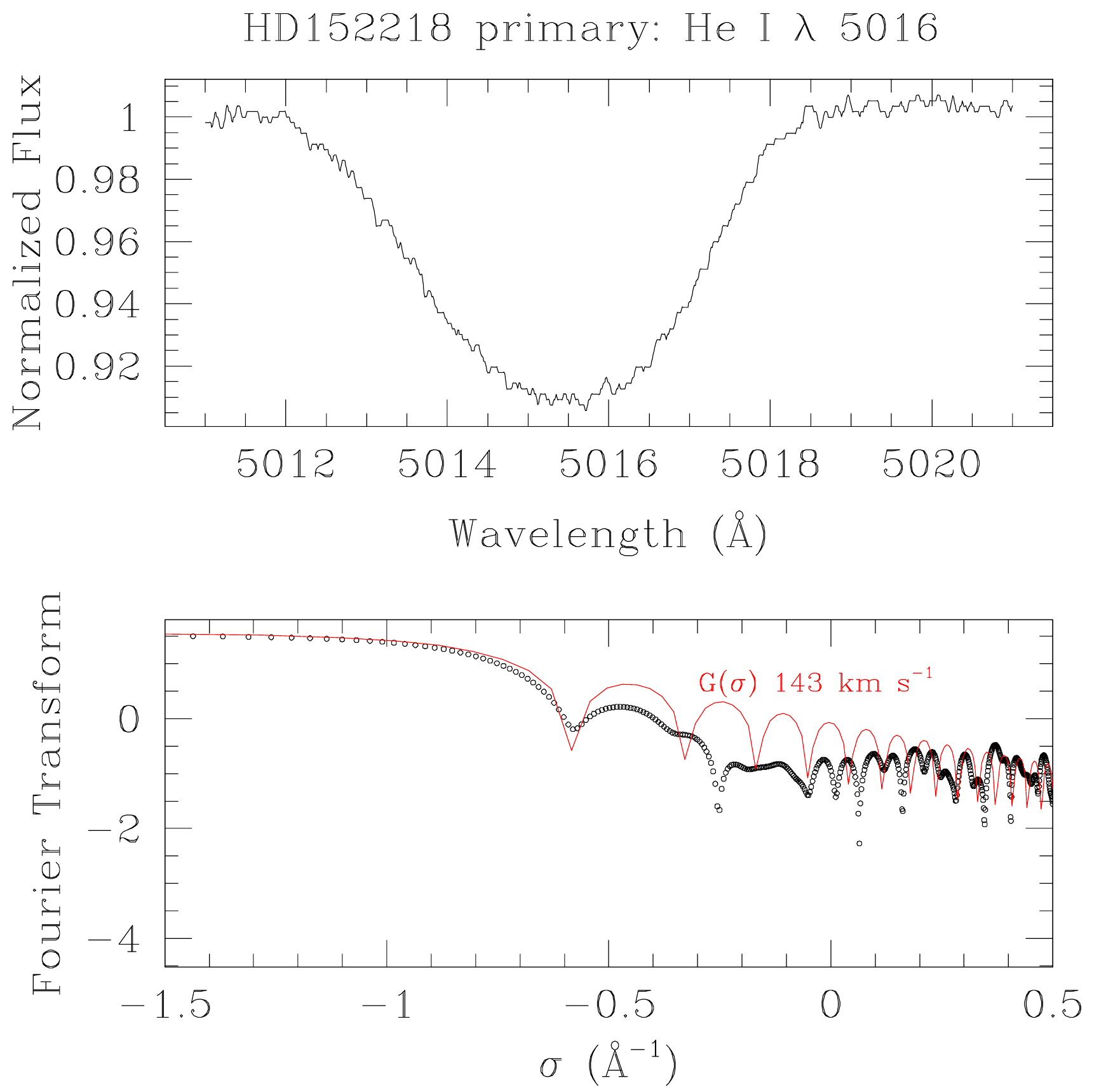}
\includegraphics*[width=0.45\textwidth,angle=0]{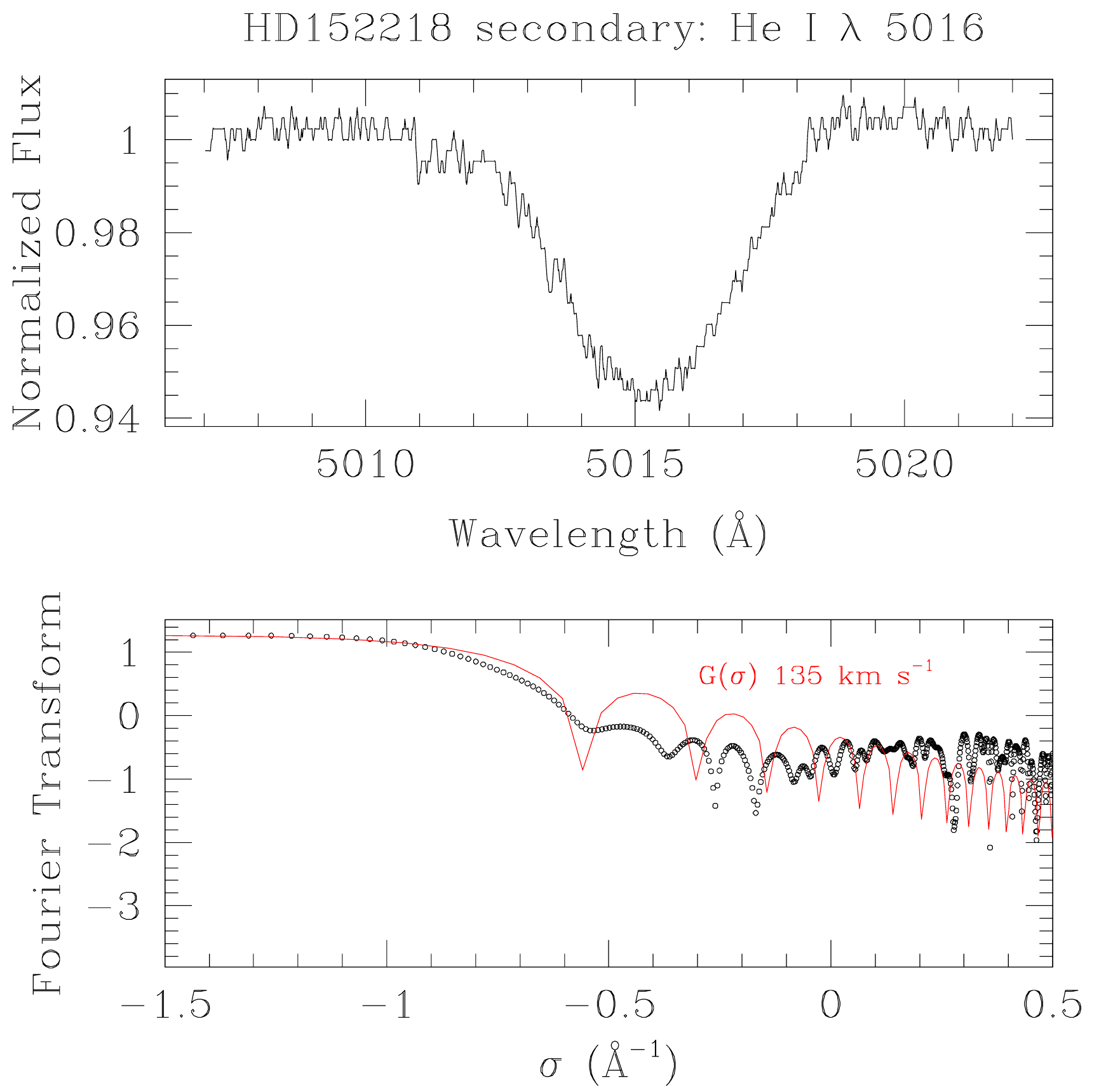}
\end{center}
\caption{He\,{\sc i} $\lambda$\,5016 line profiles as derived from the separated spectra and Fourier transform of the latter along with the best-match rotational profile for the primary (left panel) and secondary (right panel).\label{vsiniplot}}
\end{figure*}
\subsection{Spectral types and stellar parameters \label{SpT}}
We used the separated spectra of the components of HD~152218 to reassess the spectral types and to determine more precise values of the effective temperature as well as estimates of the surface gravity of the stars. 

For the spectral classification, we applied the criterion proposed by \citet{Conti} and \citet{Mathys} based on the ratio between the equivalent widths (EWs) of the He\,{\sc i} $\lambda$\,4471 and He\,{\sc ii} $\lambda$\,4542 lines. In this way, we found that $\log{W'} = \log[$EW(He\,{\sc i}\,$\lambda$\,4471)/EW(He\,{\sc ii}\,$\lambda$\,4542)$]$ amounts to $0.34 \pm 0.01$ for the primary and $0.66 \pm 0.02$ for the secondary. These values correspond to spectral types O9 and O9.7 (with O9.5 within the error bar) for the primary and secondary, respectively. To assess the luminosity classes, we compared our separated spectra with the digital atlas of \citet{WF}. The strengths of the Si\,{\sc iv} $\lambda\lambda$\,4089, 4116 and He\,{\sc ii} $\lambda$\,4686 absorptions suggest main-sequence or, especially in the case of the primary, a slightly more evolved luminosity class. We therefore classify the primary as O9\,IV, and the secondary as O9.7\,V. These results confirm the classification previously proposed by \citet{Stickland} and \citet{Sana}. 

Using the individual spectra obtained through spectral disentangling, we derived the values of the projected rotational velocities of the primary and secondary star. For this purpose, we applied the Fourier transform method \citep{Gray, SDH}  to the profiles of the He\,{\sc i} $\lambda\lambda$\,4026, 4713, 4922, and 5016 lines of both the primary and the secondary star, and to the O\,{\sc iii} $\lambda$\,5592 profile of the primary. These lines are relatively well isolated in the spectra and are therefore expected to be almost free of blends. The He\,{\sc i} $\lambda$\,4026 line might be weakly polluted by He\,{\sc ii} $\lambda$\,4026, although given the spectral types of the stars, such a pollution should be at a very low level. The results are presented in Table\,\ref{vsini} and illustrated in Fig.\,\ref{vsiniplot}. 
\begin{table}[htb]
\caption{Best-fit projected rotational velocities as derived from the separated spectra.}
\begin{center}
\begin{tabular}{c c c}
\hline\hline
Line & \multicolumn{2}{c}{$v\,\sin{i_{\rm rot}}$ (km\,s$^{-1}$)} \\
& Primary & Secondary \\
\hline
He\,{\sc i} $\lambda$\,4026 & $148 \pm 5$ & $165 \pm 5$ \\
He\,{\sc i} $\lambda$\,4713 & $138 \pm 5$ & $145 \pm 10$ \\
He\,{\sc i} $\lambda$\,4922 & $149 \pm 5$ & $120 \pm 10$ \\
He\,{\sc i} $\lambda$\,5016 & $143 \pm 5$ & $135 \pm 5$ \\
O\,{\sc iii} $\lambda$\,5592 & $145 \pm 5$ & -- \\
\hline
Mean (this work) & $144.6 \pm 4.4$ & $141.3 \pm 18.9$ \\
\hline
\citet{Howarth} & $143 \pm 14$ & $125 \pm 14$ \\
\citet{Stickland} & $152 \pm 10$ & $133 \pm 30$ \\
\hline
\end{tabular}
\end{center}
\label{vsini}
\tablefoot{The values quoted by \citet{Howarth} and \citet{Stickland} were obtained by cross-correlation techniques applied to {\it IUE} spectra.}
\end{table}

We found that the primary $v\,\sin{i_{\rm rot}}$ values are in excellent agreement for the different lines, resulting in a mean $v_1\,\sin{i_{\rm rot}}$ of $(144.6 \pm 4.4)$\,km\,s$^{-1}$, in excellent agreement with the values of \citet{Howarth} and \citet{Stickland}. For the secondary, we found a larger dispersion among the various lines, resulting in a mean $v_2\,\sin{i_{\rm rot}}$ of $(141 \pm 19)$\,km\,s$^{-1}$ , which is in reasonable agreement with the results of \citet{Howarth} and \citet{Stickland}.

To derive quantitative values of the effective temperature and the surface gravity, we used a grid of synthetic spectra generated with the CMFGEN model atmosphere code \citep{CMFGEN}. The line profiles in the synthetic spectra were broadened according to the projected rotational velocities inferred above. The relative strengths of the spectral lines in the separated spectra indicate a brightness ratio $l({\rm O9})/(l({\rm O9})+l({\rm O9.7}))$ of $0.67$ in excellent agreement with the value ($0.67 \pm 0.1$) inferred by \citet{Sana} and the corresponding UV flux ratio ($0.67$) inferred from {\it IUE} spectra \citep{Penny}. The temperatures are essentially determined by the ability to correctly reproduce the relative strengths of He\,{\sc i} and He\,{\sc ii} lines, whilst gravity is mainly determined from the wings of H\,{\sc i} Balmer lines. In many early-type binaries, especially those with small or moderate RV amplitudes, spectral disentangling has difficulties to correctly reproduce the wings of the Balmer lines, which limits the possibilities of using these wings in constraining $\log{g}$ \citep[e.g.][]{Raucq}. However, in the present case, the orbital motion is sufficiently important that the disentangling works well for these lines.

The comparison between the separated spectra and the synthetic spectra that provide the best match is shown in Fig.\,\ref{specfit}. Since we found no indication for non-solar composition, our fits were performed assuming a solar chemical composition \citep{Asplund}. 

Our CMFGEN analysis yields temperatures for the O9 and O9.7 stars of 33\,400 and 29\,900\,K, respectively. In the same way, we determine $\log{g} = 3.81$ and $3.89$ for the primary and secondary, respectively. The lower panels of Fig.\,\ref{specfit} illustrate the $\chi^2$ maps of the CMFGEN model fits in the ($T_{\rm eff}$, $\log{g}$) plane. We followed the approximation of \citet{MartinsMimes} in renormalizing the $\chi^2$ so that its minimum has a value of 1.0. The $1\sigma$ confidence contour would then be given by the $\chi^2 = 2.0$ contour \citep[see][]{MartinsMimes}. Clearly, a correlation between the two parameters exists, which makes it difficult to derive meaningful uncertainties in this case. We therefore adopted a formal error estimate of 1000\,K on $T_{\rm eff}$ and 0.15 on $\log{g}$.  

\section{Light-curve analysis \label{photometry}}
\begin{figure}
\begin{center}
\includegraphics*[width=0.5\textwidth,angle=0]{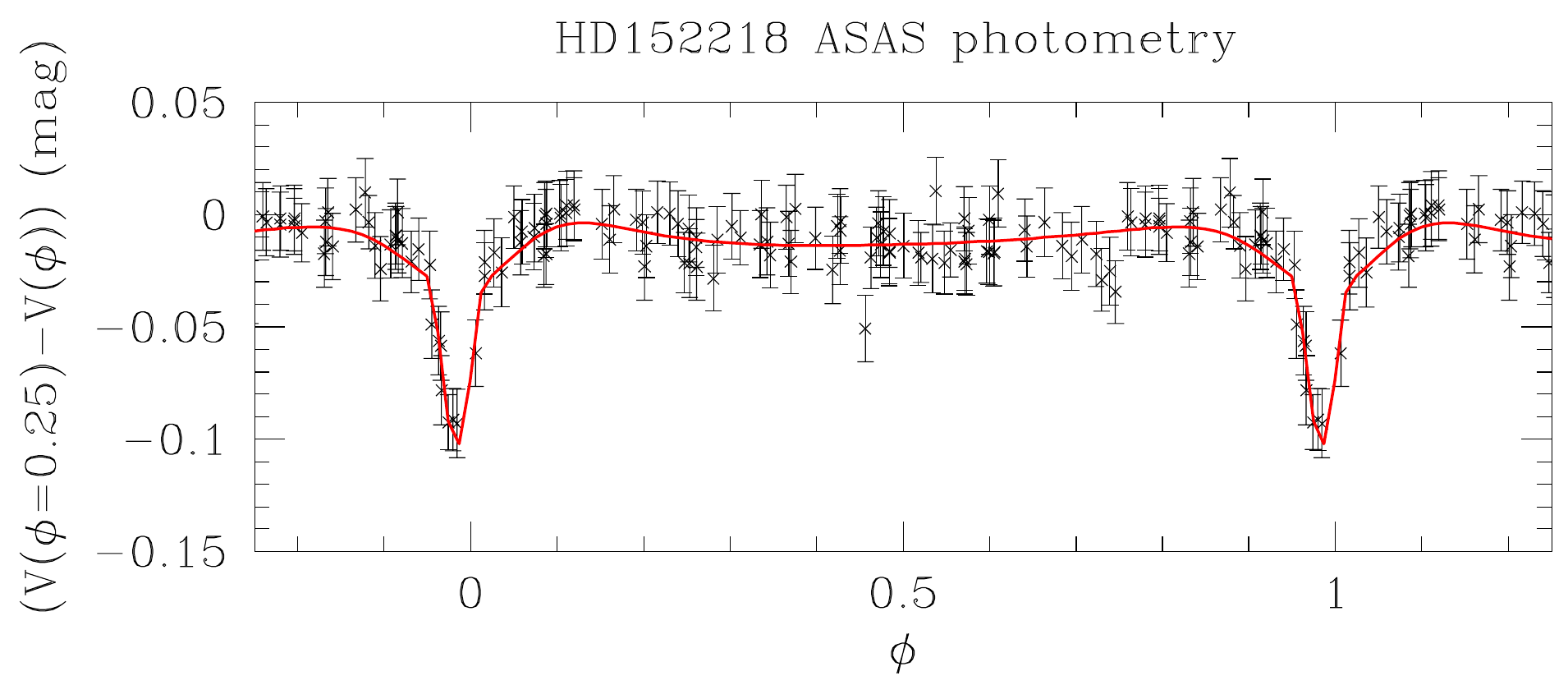}
\end{center}
\caption{Photometric light curve of HD~152218 as derived from the ASAS-3 data along with the best-fit model computed with the {\tt Nightfall} code.\label{lightcurve}}
\end{figure}

As a result of the inclination of the orbit and because of the orientation of its apses with respect to our line of sight, the light curve of HD~152218 displays only a single eclipse per orbital cycle, shortly before periastron passage when the primary star passes behind the secondary (see Fig.\,\ref{lightcurve}). 
There is only a single rather shallow eclipse. Therefore it is easy to conceive that the inversion of this light curve can be subject to degeneracies or correlations between model parameters. To limit the effect of these problems, we used the results of our spectroscopic study to constrain as many model parameters as possible. The mass-ratio, the temperatures of both stars, the orbital eccentricity, and the longitude of periastron at the epoch of the photometric observations were all frozen in our study. The only free parameters were the Roche-lobe filling factors of both stars {\it fill}(O9\,IV) and {\it fill}(O9.7\,V) and the orbital inclination ($i$). The filling factors are defined here as the fraction of the polar radius of the star under consideration over the polar radius of its instantaneous Roche lobe at periastron passage.

\begin{figure*}[htb]
\begin{center}
\includegraphics*[width=0.3\textwidth,angle=0]{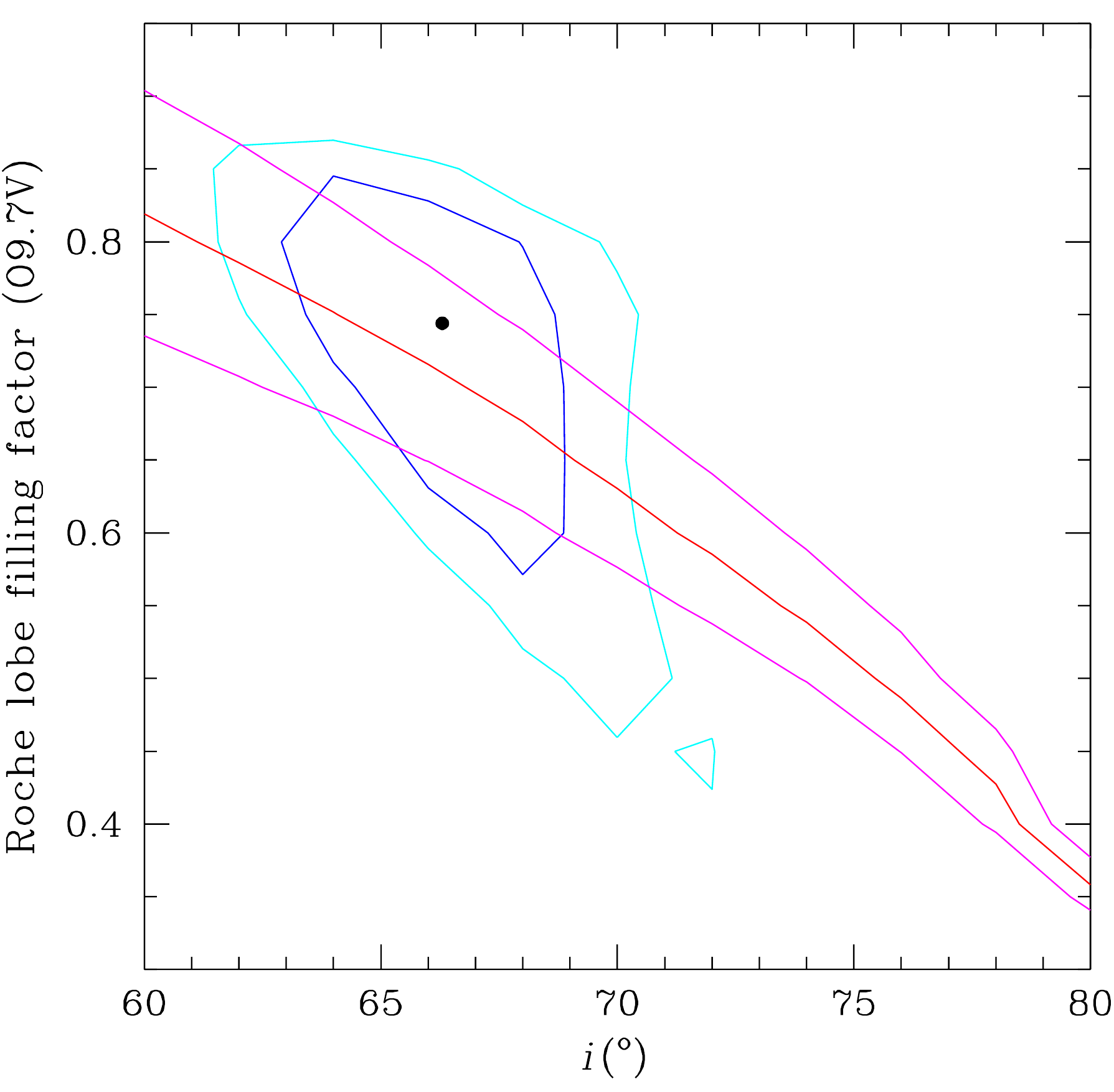}
\includegraphics*[width=0.3\textwidth,angle=0]{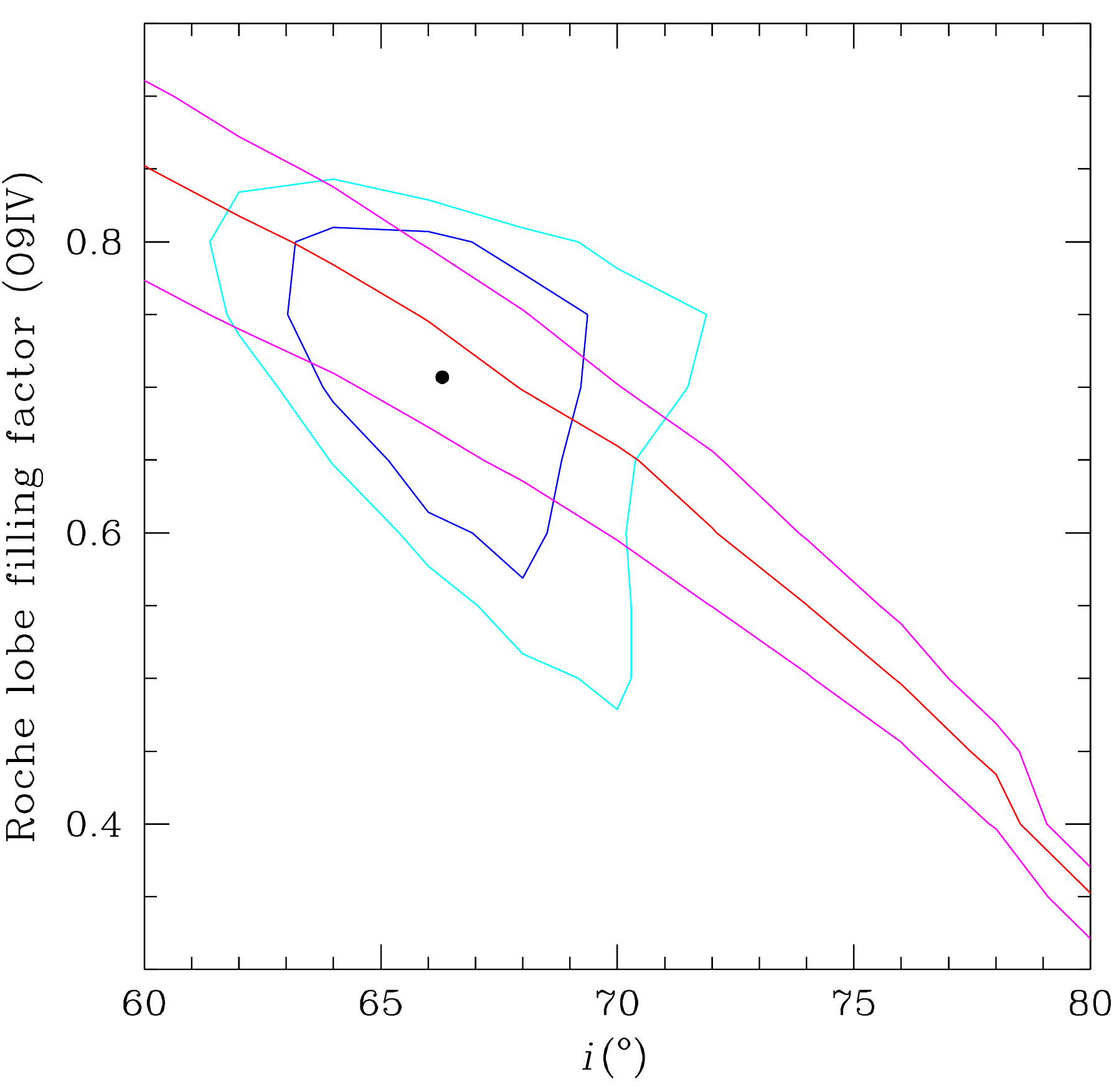}
\includegraphics*[width=0.3\textwidth,angle=0]{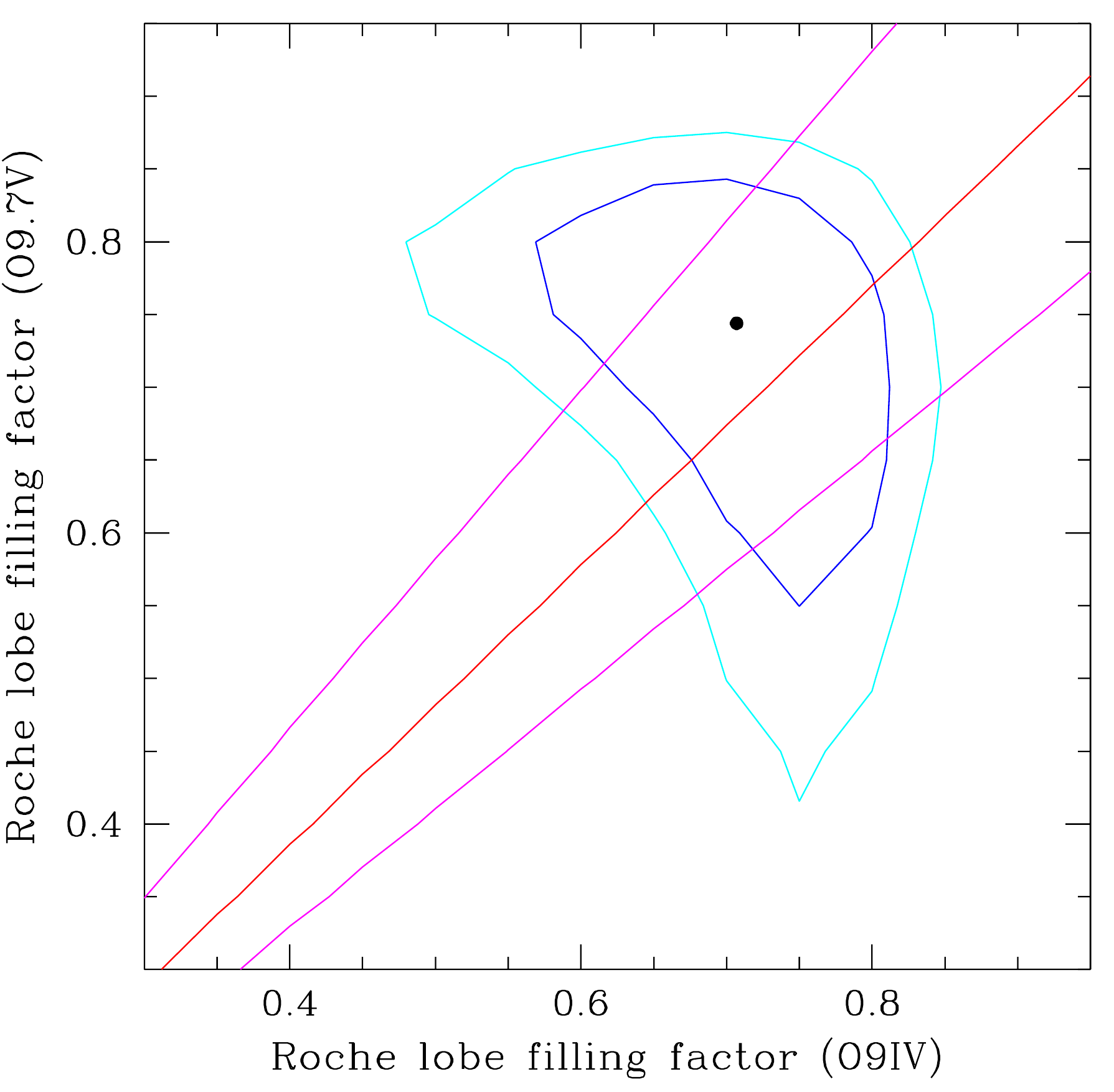}
\end{center}
\caption{Contours of 1$\sigma$ (blue) and 90\% confidence intervals (cyan) in the various parameter spaces of the light curve analysis. The red curve corresponds to a ratio of the filling factors of 1.04 (see text), whilst the magenta curves yield the corresponding 1-$\sigma$ interval, as derived from spectroscopy. The filled dot indicates the set of parameters that yields the best fit of the ASAS light curve. \label{contours}}
\end{figure*}

We then ran the {\tt Nightfall} code (version 1.86)\footnote{This code was developed and is maintained by Wichmann, Kuster, and Risse, see {\tt \tiny{ http://www.hs.uni-hamburg.de/DE/Ins/Per/Wichmann/Nightfall.html}}} to search for the combination of model parameters that yields the best fit of the data. {\tt Nightfall} handles eccentric eclipsing binaries assuming that the shape of the stars can be described by a Roche potential scaled with the instantaneous separation between the stars. 
To estimate the errors on the model parameters, we performed a systematic study of the 3D parameter space. For this purpose, we mapped the contours of $\chi^2$ over three grids that sample the projections of the 3D parameter space onto the planes formed by the combinations of two out of the three parameters {\it fill}(O9\,IV), {\it fill}(O9.7\,V), $i$. The resulting contours for $\Delta \chi^2 = 2.30$ and $4.61$ ($1\sigma$ and 90\% confidence intervals for two degrees of freedom) are displayed in Fig.\,\ref{contours}. 

However, not all combinations of parameters that are acceptable as far as the fit of the light curve is concerned agree with the spectroscopic results. From spectroscopy, we have inferred an optical brightness ratio that leads to a ratio of bolometric luminosities of $L_{\rm bol}$(O9\,IV)$/L_{\rm bol}$(O9.7\,V)$ = 2.17 \pm 0.75$. With the temperatures of the stars, this ratio translates into $$R_*({\rm O9\,IV})/R_*({\rm O9.7\,V}) = 1.18 \pm 0.20.$$ Accounting for the ratio of the radii of the primary and secondary Roche lobes, we then inferred {\it fill}(O9\,IV)/{\it fill}(O9.7\,V) $= 1.04 \pm 0.18$. The curves corresponding to this condition are also shown in Fig.\,\ref{contours}. Their intersections with the $\chi^2$ contours for $1\sigma$ confidence intervals were used to establish the error bars on the parameters (see Table\,\ref{absolute}). Whilst the orbital inclination is rather well constrained, the relative uncertainties on the filling factors and hence on the stellar radii are quite large. The latter situation mainly stems from a slight anti-correlation between the two parameters that leads to elongated and inclined confidence contours in the {\it fill}(O9.7\,V) versus {\it fill}(O9\,IV) plane (see Fig.\,\ref{contours}).
 
\begin{table}
\caption{Physical parameters of the components of HD~152218 as derived from our CMFGEN and photometric analyses.}
\begin{center}
\begin{tabular}{c c c c}
\hline\hline
Parameter & Primary & Secondary & Reference\\
\hline
\vspace*{-3mm}\\
{\it fill} & $0.71^{+0.10}_{-0.09}$ & $0.74^{+0.09}_{-0.15}$ & Sect.\,\ref{photometry}\\
\vspace*{-3mm}\\
$R_*$\,(R$_{\odot}$) & $8.4^{+1.1}_{-1.0}$ & $7.8^{+1.0}_{-1.6}$ & Sect.\,\ref{photometry} \& Table\,\ref{bestfit}\\
\vspace*{-3mm}\\
$i$\,($^{\circ}$) & \multicolumn{2}{c}{$66.3^{+3.0}_{-3.3}$} & Sect.\,\ref{photometry} \\
\vspace*{-3mm}\\
$m$\,(M$_{\odot}$) & $19.8 \pm 1.5$ & $15.0 \pm 1.1$  & Sect.\,\ref{photometry} \& Table\,\ref{bestfit}\\
\vspace*{-3mm}\\
$T_{\rm eff}$\,(K) & $33\,400 \pm 1000$ & $29\,900 \pm 1000$ & Sect.\,\ref{SpT} \\
\vspace*{-3mm}\\
$\log{g}_{\rm spec}$ & $3.81 \pm 0.15$ & $3.89 \pm 0.15$ & Sect.\,\ref{SpT} \\
\vspace*{-3mm}\\
$\log{g}_{\rm photom }$ & $3.89^{+.11}_{-.12}$ & $3.83^{+.20}_{-.12}$ & Sect.\,\ref{photometry} \& Table\,\ref{bestfit}\\
\vspace*{-3mm}\\
\hline
\end{tabular}
\end{center}
\label{absolute}
\end{table}

From these considerations we can then estimate absolute radii and masses of the components of HD~152218 as given in Table\,\ref{absolute}. In this table we also compare the values of $\log{g}$ obtained from our CMFGEN analysis with those that we obtain using the absolute masses and radii as determined from the analysis of the light curve. Within the error bars, both estimates agree.

Assuming that the rotational axes of the stars are aligned with the orbital axis, we can use our best-fit value of the inclination and stellar radii and combine them with the projected rotational velocities from Table\,\ref{vsini}. In this way, we derive rotational periods of $2.69^{+0.37}_{-0.37}$\,days for the primary and $2.79^{+0.74}_{-0.52}$\,days for the secondary. Within the error bars, both stars thus appear to have their rotations nearly synchronized with each other. The rotational periods we derived are shorter by about a factor two than the orbital period. When we compare the rotational angular velocities with the instantaneous orbital angular velocity at periastron, we find ratios of $1.13^{+0.12}_{-0.12}$ and $1.08^{+0.24}_{-0.17}$ for the primary and secondary, respectively. We therefore conclude that pseudo-synchronization \citep{Hut} has been achieved in the system, or is on the verge of being achieved. These properties make HD~152218 an interesting target to search for tidally induced oscillations \citep[][and references therein]{Moreno,Palate}.

\section{Apsidal motion \label{theory}}
The observationally derived rate of apsidal motion can be compared with the rates predicted from theoretical models. The rate of apsidal motion in a close binary system can be directly related to the internal structure constants $k_2$ of both stars \citep[e.g.][]{Shakura}. The internal structure constants are obtained from
\begin{equation}
k_2 = \frac{3 - \eta_2(R_*)}{4 + 2\,\eta_2(R_*)}
,\end{equation}
where $\eta_2(R_*)$ is the solution of the Clairaut-Radau differential equation
\begin{equation}
r\,\frac{d\,\eta_2(r)}{dr} + \eta_2^2(r) - \eta_2(r) + 6\,\frac{\rho(r)}{<\rho>(r)}\,(\eta_2(r) +1) - 6 = 0
\label{Radau}
,\end{equation}
evaluated at the stellar surface. In Eq.\,\ref{Radau}, $r$ is the radial distance from the centre of the star, $\rho(r)$ is the local density, and $<\rho>(r)$ is the mean density inside a sphere of radius $r$ \citep[e.g.][]{Sterne,Schmitt}. Solving the Clairaut-Radau equation applied to the density structure predicted by stellar evolution models thus yields the value of $k_2$ for a given stellar mass and age. 

To this aim, we have designed a FORTRAN code using a fourth-order Runge-Kutta method with step doubling \citep{RK}. To validate our code, we adopted the same method as \citet{IB}, considering a series of polytropic models. For this purpose, we solved the Lane-Emden differential equation for different values of the polytropic index $n$ and then applied our Clairaut-Radau integration routine to the resulting density profiles. We then compared our values of $k_2$ to those listed by \citet{BO} and found an excellent agreement with relative differences of less than $10^{-5}$.\\

After the validation of our code, we applied it to a series of stellar structure models computed with the Code Li\'egeois d'Evolution Stellaire \citep[CLES,][]{Scuflaire} for stars of 15 and 20\,M$_{\odot}$. Internal mixing was either restricted to the convective core or included a core overshooting of 0.2 times the pressure height scale ($\alpha_{ov} = 0.2$). The chemical composition of the stars was assumed to be solar adopting the abundances of \citet{Asplund}. Mass-loss was included according to the prescription of \citet{Vink}. We tested both a solar metallicity ($Z = 0.014$) and a slightly increased metallicity of $Z = 0.020$. 

Table\,\ref{k2CLES} lists the model stellar radii and corresponding $k_2$ values as a function of stellar age. The decrease in $k_2$ values as the stars evolve is evident. At a given age, overshooting has a moderate effect on both the stellar radius, which decreases compared to the models without overshooting, and the value of $k_2$, which decreases for the 20\,M$_{\odot}$ models, but increases for the 15\,M$_{\odot}$ models. The effect of an enhanced metallicity is to increase the stellar radii and simultaneously decrease the values of $k_2$. 

\begin{table*}
\begin{center}
\caption{Evolution of the internal structure constant $k_2$  and the radius of 15\,M$_{\odot}$ and 20\,M$_{\odot}$ CLES model stars as a function of age, overshooting parameter, and metallicity. \label{k2CLES}}
\begin{tabular}{|c | r c | r c | r c | r c| r c | r c|} 
\hline\hline
Age  & \multicolumn{6}{c|}{20\,M$_{\odot}$} & \multicolumn{6}{c|}{15\,M$_{\odot}$} \\
\cline{2-13}
(Myr)& \multicolumn{2}{c|}{$\alpha_{ov}=0.0$} & \multicolumn{2}{c|}{$\alpha_{ov}=0.2$} & \multicolumn{2}{c|}{$\alpha_{ov}=0.0$} & \multicolumn{2}{c|}{$\alpha_{ov}=0.0$} & \multicolumn{2}{c|}{$\alpha_{ov}=0.2$} & \multicolumn{2}{c|}{$\alpha=0.0$} \\
& \multicolumn{2}{c|}{$Z=0.014$} & \multicolumn{2}{c|}{$Z=0.014$} & \multicolumn{2}{c|}{$Z=0.020$} & \multicolumn{2}{c|}{$Z=0.014$} & \multicolumn{2}{c|}{$Z=0.014$} & \multicolumn{2}{c|}{$Z=0.020$} \\
   & $R_*$\,(R$_{\odot}$) & $k_2$ & $R_*$\,(R$_{\odot}$) & $k_2$ & $R_*$\,(R$_{\odot}$) & $k_2$ & $R_*$\,(R$_{\odot}$) & $k_2$ & $R_*$\,(R$_{\odot}$) & $k_2$ & $R_*$\,(R$_{\odot}$) & $k_2$ \\
\hline  
3.3  &  6.834 & .010115 &  6.778 & .010025 & 7.131 & .009241 & 5.420 & .011154 & 5.366 & .011196 & 5.615 & .010386 \\
3.8  &  7.142 & .009187 &  7.051 & .009127 & 7.441 & .008414 & 5.548 & .010632 & 5.485 & .010677 & 5.747 & .009905 \\
4.3  &  7.481 & .008326 &  7.370 & .008240 & 7.810 & .007587 & 5.690 & .010104 & 5.612 & .010157 & 5.890 & .009428 \\
4.8  &  7.923 & .007403 &  7.753 & .007350 & 8.258 & .006757 & 5.842 & .009586 & 5.749 & .009631 & 6.046 & .008951 \\
5.3  &  8.447 & .006528 &  8.224 & .006468 & 8.803 & .005946 & 6.012 & .009065 & 5.901 & .009113 & 6.218 & .008472 \\
5.8  &  9.110 & .005663 &  8.806 & .005602 & 9.495 & .005143 & 6.206 & .008532 & 6.069 & .008589 & 6.410 & .007989 \\
6.3  & 10.010 & .004788 &  9.559 & .004749 & 10.421 & .004337 & 6.426 & .008002 & 6.256 & .008059 & 6.628 & .007503 \\
6.8  & 11.258 & .003921 & 10.604 & .003894 & 11.720 & .003531 & 6.677 & .007467 & 6.470 & .007529 & 6.877 & .007015 \\
7.3  & 12.789 & .003128 & 12.139 & .003046 & 13.425 & .002776 & 6.963 & .006937 & 6.715 & .006994 & 7.158 & .006529 \\
\hline
\end{tabular}
\end{center}
\end{table*}
 
\begin{figure}[htb]
\begin{center}
\includegraphics*[width=0.45\textwidth,angle=0]{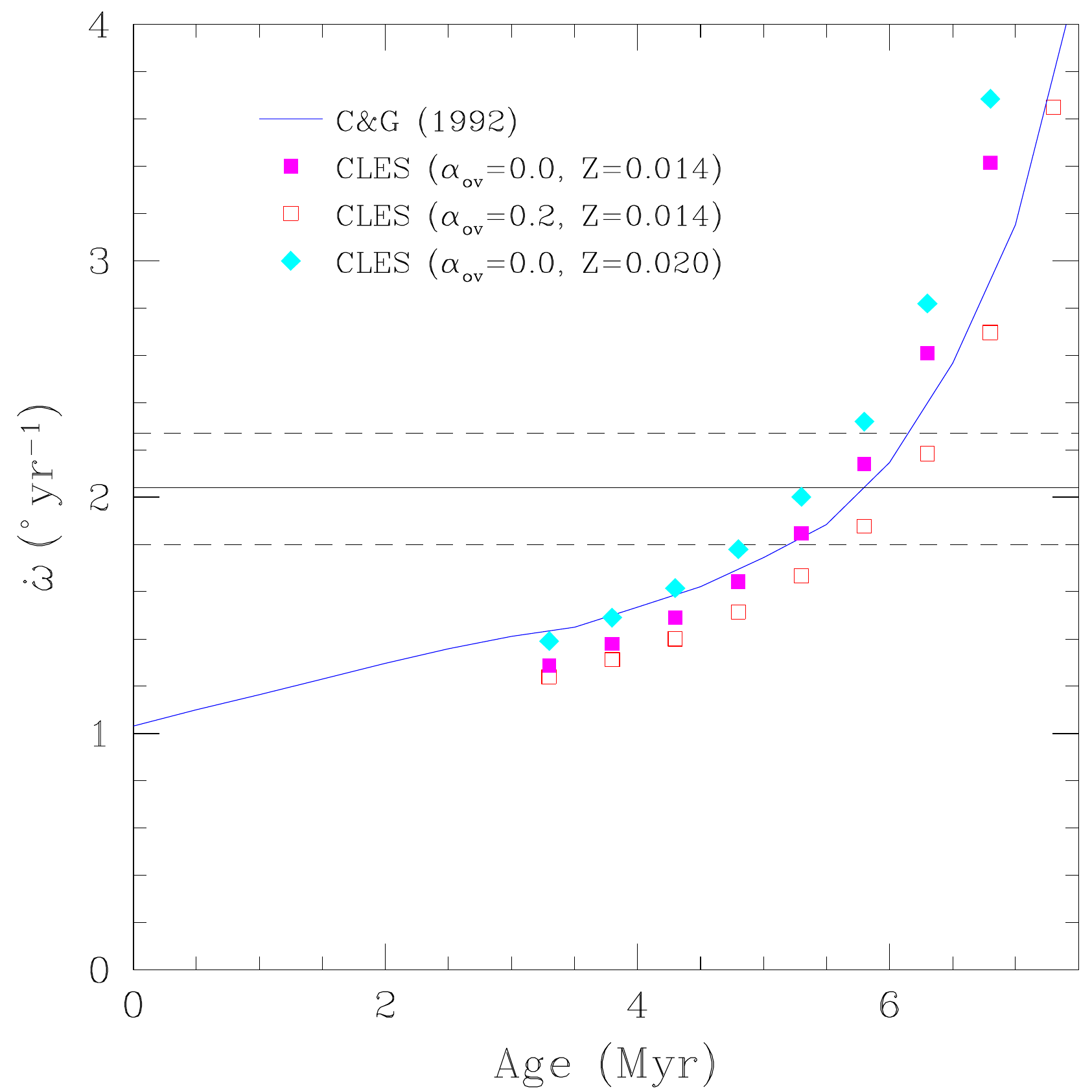}
\end{center}
\caption{Comparison between the observed value of $\dot{\omega}$ (solid horizontal line) with its errors (dashed horizontal lines) and the values evaluated from stellar structure models as a function of stellar age. The solid curve yields the results obtained by interpolating the models of \citet{CG92}. The filled squares correspond to the results obtained through the resolution of the Clairaut-Radau equation applied to the CLES models for a 20\,M$_{\odot}$ primary and a 15\,M$_{\odot}$ secondary. The open squares yield the results for CLES models assuming an overshooting with $\alpha_{ov} = 0.2$. The filled diamonds stand for the CLES models with a slightly higher metallicity.\label{omegadot}}
\end{figure}

We then computed the theoretical rates of apsidal motion that are due to the linear part of the quadrupole moment for the binary components and the relativistic contribution using the following equation adapted from \citet{Shakura}:
\begin{eqnarray}
\dot{\omega} & = & \left[f(e)\left(\frac{k_{2,1}}{q}\left(\frac{R_{*,1}}{a}\right)^5+k_{2,2}\,q\,\left(\frac{R_{*,2}}{a}\right)^5\right)\right. \nonumber\\
& + & g(e)\left(k_{2,1}\left(\frac{R_{*,1}}{a}\right)^5\left(\frac{P_{orb}}{P_{rot,1}}\right)^2\frac{1+q}{q}\right. \nonumber \\
& + & \left.\left.k_{2,2}\left(\frac{R_{*,2}}{a}\right)^5\left(\frac{P_{orb}}{P_{rot,2}}\right)^2\,(1+q)\right)\right]\frac{2\,\pi}{P_{orb}} + \dot{\omega}_{\rm GR} \label{omegadotShakura}
,\end{eqnarray}
where 
\begin{equation}
f(e)  =  15\,\frac{1+\frac{3\,e^2}{2}+\frac{e^4}{8}}{(1-e^2)^5} 
\end{equation}
and
\begin{equation}
g(e)  =  \frac{1}{(1 - e^2)^2} 
.\end{equation}

$\dot{\omega}_{\rm GR}$ stands for the rate of apsidal motion due to general relativity \citep{Gimenez}. Indices 1 and 2 refer to the primary and secondary star, respectively, and $q = \frac{m_1}{m_2}$. As we are interested in determining how the theoretical rate of apsidal motion evolves with age, the stellar radii used in Eq.\,\ref{omegadotShakura} were taken from the evolutionary models rather than from the photometric solution. The rotational periods were computed from the projected equatorial rotational velocities (see Table\,\ref{vsini}), assuming the rotation axes to be aligned with the normal to the orbital plane and adopting again the radii of the stars as predicted by the evolutionary models. As noted in Sect.\,\ref{photometry}, HD~152218 is currently most likely in a pseudo-synchronization state. As an alternative to our above approach to the calculation of the rotational periods, we might therefore assume that both stars have reached pseudo-synchronization early on. Using this assumption instead of the values of $v\,\sin{i_{\rm rot}}$ does not change the predicted values of $\dot{\omega}$ in a significant way, at least not as far as the range of ages is concerned where the predicted matches the observed value.

The stellar radii play a key role in determining the value of $\dot{\omega}$ because the $f(e)$ term of the rate of apsidal motion depends on $(R_*/a)^5$, whilst the $g(e)$ term still depends on $(R/a)^3$ (see Eq.\,\ref{omegadotShakura}). Therefore, the decrease of the stellar radii for the models with overshooting more than compensates for the changes of $k_2$, resulting in lower predicted values of $\dot{\omega}$. In the same way, the larger radii of the higher $Z$ models, compared to models with $Z = 0.014$ of the same age, lead to higher theoretical values of $\dot{\omega}$.

Figure\,\ref{omegadot} illustrates the comparison between our observationally determined value of $\dot{\omega}$ and the theoretical predictions based on the CLES models (Table\,\ref{k2CLES}) for primary and secondary masses of 20 and 15\,M$_{\odot}$ and adopting the eccentricity and orbital period from Table\,\ref{bestfit}. The solid curve in Fig.\,\ref{omegadot} furthermore illustrates the results of an interpolation of the \citet{CG92} models (again for primary and secondary masses of 20 and 15\,M$_{\odot}$), which assume $\alpha_{ov} = 0.2$ and $Z=0.020$. In each case, we include a relativistic contribution of $\dot{\omega}_{\rm GR} = 0.13^{\circ}$\,yr$^{-1}$ evaluated according to \citet{Shakura} and \citet{Gimenez}. 

The comparison between the model predictions and the observations yields an estimated age of the HD~152218 binary system in the range between 5.2 and 6.4\,Myr, accounting for the uncertainties on $\dot{\omega}$ and the uncertainties due to the different prescriptions for $\alpha_{ov}$, and assuming $Z=0.014$. When instead we were to assume $Z=0.020$, our age estimate would be reduced by 0.3\,Myr. 

Finally, we have also tested models with an enhanced helium abundance corresponding to $X = 0.66$ instead of $X = 0.70$ adopted above. For a given age, these models have larger radii that lead to higher predicted values of $\dot{\omega}$ and thus an age between 4.5 and 5.5\,Myr. We caution, however, that the analysis of seven B-type stars by \citet{Mathys2} resulted in $\log{\epsilon_{\rm He}} = 10.81 \pm 0.20$, whilst an earlier study of five B-stars by \citet{Keenan} yielded $\log{\epsilon_{\rm He}} = 10.92 \pm 0.11$. Both results are compatible with the solar He abundance \citep[$\log{\epsilon_{\rm He}} = 10.93 \pm 0.01$,][]{Asplund}. Furthermore, our separated spectra of HD~152218 do not provide evidence of an enhanced helium abundance. This means that observational data support our $X = 0.70$ results.  

Assuming the rotational axes to be aligned with the direction perpendicular to the plane of the orbit, \citet{Benvenuto} have shown that the rate of apsidal motion (Eq.\,\ref{omegadotShakura}) can be expressed as a function of the primary mass as well as of several parameters that can be determined observationally (mass ratio, projected rotational velocities, eccentricity, etc.) or theoretically (internal structure constants and stellar radii). A simplified formalism based on the same idea was proposed by \citet{Jeffery}. We note that 
\begin{equation}
\left(\frac{R_{*,j}}{a}\right)^5 = \left(\frac{2\,\pi}{P_{orb}}\right)^{10/3}\,G^{-5/3}\,\left(\frac{q}{1 + q}\right)^{5/3}\,R_{*,j}^5\,m_1^{-5/3} 
\end{equation}
\begin{equation}
\left(\frac{R_{*,j}}{a}\right)^5\,\left(\frac{P_{orb}}{P_{rot,j}}\right)^2 =  \left(\frac{v_j\,\sin{i}}{a\,\sin{i}}\right)^2\,\frac{R_{*,j}^3\,q}{G\,m_1\,(1 + q)}
\end{equation}
and 
\begin{equation}
\dot{\omega}_{\rm GR} = \frac{3\,G^{2/3}}{c^2\,(1 - e^2)}\,\left(\frac{2\,\pi}{P_{orb}}\right)^{5/3}\,\left(m_1\,\frac{1 + q}{q}\right)^{2/3}
,\end{equation}
which lead to 
\begin{eqnarray}
\dot{\omega} & = & \left[f(e)\left(\frac{2\,\pi}{P_{orb}}\right)^{10/3} \left(\frac{q}{G\,m_1\,(1 + q)}\right)^{5/3}\,\left(\frac{k_{2,1}}{q}\,R_{*,1}^5+k_{2,2}\,q\,R_{*,2}^5\right)\right. \nonumber\\
& + & \left.g(e)\left(k_{2,1}\left(\frac{v_1\,\sin{i}}{a\,\sin{i}}\right)^2\,\frac{R_{*,1}^3}{G\,m_1} + k_{2,2}\left(\frac{v_2\,\sin{i}}{a\,\sin{i}}\right)^2\,\frac{R_{*,2}^3\,q}{G\,m_1}\right)\right]\frac{2\,\pi}{P_{orb}} \nonumber \\
& + &  \frac{3\,G^{2/3}}{c^2\,(1 - e^2)}\,\left(\frac{2\,\pi}{P_{orb}}\right)^{5/3}\,\left(m_1\,\frac{1 + q}{q}\right)^{2/3} \label{omegadotBenvenuto}
.\end{eqnarray}
\citet{Benvenuto} used this formulation of apsidal motion to constrain the masses of the components of the non-eclipsing O3\,V + O8\,V binary HD~93205 by comparing the observed rate of apsidal motion against the predictions from a grid of theoretical models \citep[see also][for other applications of this method]{Ferrero,Pablo}. Whilst this method for constraining the masses is by construction highly model dependent and sensitive to the assumed age, it is interesting to apply it to HD~152218.

\begin{figure}[htb]
\begin{center}
\includegraphics*[width=0.45\textwidth,angle=0]{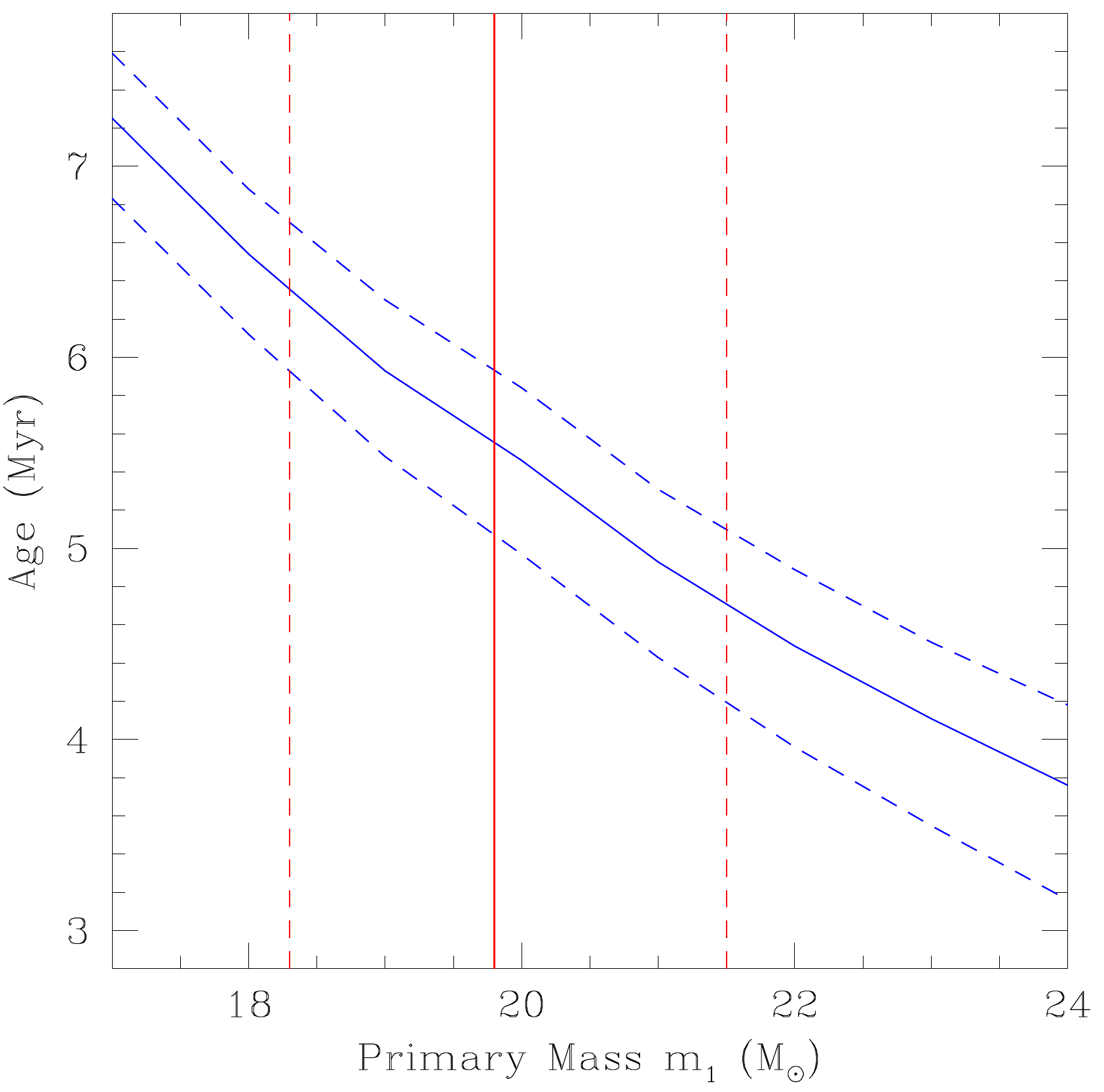}
\end{center}
\caption{Relation between the age of HD~152218 and the mass of the primary star. The blue solid curve corresponds to the combination of age and primary mass in our CLES models (assuming $Z=0.014$ and $\alpha = 0.0$) that reproduce exactly the observed rate of apsidal motion through Eq.\,\ref{omegadotBenvenuto}. The blue dashed curves correspond to the 1-$\sigma$ errors on the observed $\dot{\omega}$. The red vertical lines correspond to the absolute mass of the primary determined from our orbital and photometric solution (Table\,\ref{absolute}) and its errors.\label{massage}}
\end{figure}

For this purpose, we have computed a grid of CLES evolutionary models with $\alpha = 0.0$ and $Z=0.014$ for masses between 11 and 24\,M$_{\odot}$. Using our newly determined binary parameters ($q$, $e$, $P_{orb}$, $a\,\sin{i}$) and projected rotational velocities, we then determined the age of the system that allows us to reproduce the observed rate of apsidal motion for a given value of $m_1$. The results are shown in Fig.\,\ref{massage}.

Using the absolute mass of the primary and its uncertainty (see Table\,\ref{absolute}), we can then estimate an age of $5.6^{+1.1}_{-1.4}$\,Myr, broadly confirming the results obtained above. 

\section{Discussion \label{discussion}}
The age of the early-type stellar population of the open cluster NGC~6231 has been estimated using various techniques, and an overview of the relevant literature can be found in \citet{Reipurth}. Age determinations of young open clusters mostly rely upon isochrone fitting to the population of early-type stars. They therefore obviously depend on the ingredients of the evolutionary models (e.g.\ opacities, metallicity, inclusion or not of rotation, etc.) and the accuracy of calibrations inferred from stellar atmosphere models (e.g.\ bolometric corrections, effective temperatures, etc.). These methods are also biased by the inclusion of binary systems that are not recognized as such. Finally, as revealed by studies of the populations of pre-main sequence stars, low-mass star formation in young open clusters proceeds over at least about 10\,Myr, and it is rather unlikely that all massive cluster members formed simultaneously. For NGC~6231, \citet{Raboud} and \citet{Sung} reported on an age spread among the low-mass pre-main sequence stars of at least 7\,Myr \citep{Raboud} or 12\,Myr \citep{Sung}, which was subsequently confirmed by \citet{PMS}. We therefore expect a spread of ages and not a single value. With these caveats in mind, we briefly consider the existing age determinations of the massive stars in NGC~6231. 

\citet{vanGenderen} compared {\it V\,B\,L\,U\,W} photometry of 9 O-type members of the Sco\,OB1 association with isochrones and inferred an average age of $3.6 \pm 0.6$\,Myr. However, only four stars in their sample are members of NGC~6231. Moreover, three of them (HD~152218, HD~152247, and HD~152248) are binaries, whilst the fourth star (HDE~326331) displays a variable radial velocity \citep{NGC6231}. 

\citet{Perry} compiled photometric data from the literature. Most of their stars had masses between 10 and 25\,M$_{\odot}$ and were consistent with $\log({\rm age}) = 6.9 \pm 0.2$ (i.e.\ an age of $7.9^{+4.6}_{-2.9}$\,Myr), although they noted that a few more massive stars were apparently younger.

\citet{Raboud} studied Geneva photometry of the NGC~6231 cluster and fitted isochrones to the upper main-sequence, inferring an age of $3.8 \pm 0.6$\,Myr. These authors explicitly excluded known binary stars from their analysis. Still, the census of massive binaries in the cluster was not complete at that time, and it might well be that their sample still included some binaries.

\citet{Sung} performed isochrone fitting to {\it U\,B\,V\,R\,I\,H$\alpha$} photometry and inferred an age of between 2.5 and 4.0\,Myr for the massive star population.

Finally, \citet{Baume} studied {\it U\,B\,V\,I} photometry. These authors emphasized the difficulties that the large number of binaries introduces in the age determination. They estimated an age of between 3 and 5\,Myr. 

In summary, whilst there is no consensus on the age of NGC~6231, most determinations, with the exception of the study of \citet{Perry}, favour a value near $3.8 \pm 0.5$\,Myr. We stress that these literature results assumed $Z=0.020,$ although \citet{Baume} drew attention to the work of \citet{Kilian}, who analysed the chemical composition for a sample of ten B-type stars in NGC~6231 and found their CNO mass fraction to be subsolar (between 0.7 and 1.3\% compared to the then-accepted value of 1.8\% for the Sun).\\ 

As our newly determined value of the age of HD~152218 is significantly higher than the estimates above, we need to ask ourselves whether our value could be biased in some way. 

Since we used single-star evolution models in our analysis, our results rely upon the assumption that HD~152218 has not yet experienced any mass and angular momentum exchange through Roche-lobe overflow (RLOF). If the system were in a post-RLOF evolutionary stage, the exchange of mass between the stars could have led to an apparent rejuvenation of the mass gainer \citep{DT} and would in any case affect our determination of the age. However, given the properties of the system, the assumption that the system has not yet experienced any RLOF interaction seems quite reasonable. In fact, contrary to well-established post-RLOF systems \citep[e.g.][]{Linder,LZCep,Raucq}, the spectra of the components of HD~152218 do not show evidence of enhanced nitrogen abundance or asynchronous rotation. Furthermore, the orbital eccentricity is quite high for a short-period binary, and the Roche-lobe filling factors at periastron inferred from the photometric light curve (see Sect.\,\ref{photometry}) are well below unity. Therefore, it seems very unlikely that HD~152218 has experienced binary interaction in the past. 

In Sect.\,\ref{theory} we investigated the effect of overshooting on the values of $k_2$ and $\dot{\omega}$. Whilst the assumptions on internal mixing clearly do have an effect on the age determination, this is rather moderate, as illustrated by the relatively good agreement between different models in Fig.\,\ref{omegadot}. An additional effect that might influence the predictions of the theoretical models, but is not accounted for in our models, is rotational mixing. Rotational mixing in massive stars changes their position in the Hertzsprung-Russell diagram \citep[e.g.][]{MM}. However, the rotational velocities that we have derived (see Table\,\ref{vsini}) are rather moderate for massive stars. Moreover, by comparing evolutionary tracks for a non-rotating 20\,M$_{\odot}$ star and a star of the same mass rotating at $\sim 132$\,km\,s$^{-1}$ \citep[see Fig.\,7 of][]{MM}, we note that there is almost no difference in luminosity (hence in radii) between the models when the stars have effective temperatures of about 33\,400\,K ($\log{T_{\rm eff}} \simeq 4.52$). It therefore appears unlikely that the neglect of rotation in our models could lead to a significant bias on the age determination. \citet{Claret} investigated the effect of rotation on the internal structure constants. He found that for stars that are not too heavily distorted, rotation induces a reduction of the theoretical $k_2$ given by
\begin{equation}
\log{k_2^{\rm rot}} = \log{k_2^{\Omega = 0}} - 0.87\,\frac{2\,\Omega_s^2\,R_*^3}{3\,G\,m_*}
,\end{equation}
where $k_2^{\rm rot}$ and $k_2^{\Omega = 0}$ refer to the internal structure constants with and without rotation, and $\Omega_s$ is the angular rotational velocity at the stellar surface. Applying this recipe to our case leads to a reduction of the internal structure constants by $\sim 8$ and $\sim 7$\% for the primary and secondary star, respectively. Neglecting any effect of rotation on the stellar radii (see above), we then expect a reduction of the theoretical rate of apsidal motion by the same $\sim 8$\%. Therefore, accounting for rotation leads to even higher values of the ages.

Finally, we made the assumption that the rotational axes are aligned with the normal to the orbital plane. \citet{Shakura} presented a formalism (his Eq.\ 3) that allows us to deal with situations where this assumption is not fulfilled. Assuming that the rotation periods of the two binary components are known, the angle-dependent part of the rotation term of $\dot{\omega}$ takes the form \citep{Shakura}
\begin{equation}
F(i,i_{rot},\gamma) = -\frac{\cos{\gamma}\,(\cos{\gamma}-\cos{i_{\rm rot}}\,\cos{i})}{\sin^2{i}} - \frac{1 - 5\,\cos^2{\gamma}}{2}
\label{e1}
,\end{equation}
where $i$ denotes the inclination of the orbital plane with respect to the observer, $i_{\rm rot}$ is the angle of the rotation axis with respect to the observer, and $\gamma$ is the misalignment angle between the normal to the orbit and the rotation axis. For alignment we have $\gamma = 0$ and $i_{\rm rot} = i$, thus $F(i,i,0) = 1$.  
We adopt a Cartesian coordinate system with the $z$-axis pointing at the observer and the $y$-axis chosen such that the vector normal to the orbital plane is located in the $(y,z)$ plane. To uniquely specify the rotation axis in space we need (in addition to the already introduced angle $\gamma$ ) an azimuthal angle $\theta$ that measures the precession angle of the stellar spin axis around the normal to the orbital plane. With the convention that $\theta = 0$ yields a spin vector in the $(y,z)$ plane, this yields the following relation between the angles $\theta$, $\gamma$, $i$, and $i_{\rm rot}$:
\begin{equation}
\cos{i_{\rm rot}} = \cos{i}\,\cos{\gamma} + \sin{i}\,\sin{\gamma}\,\cos{\theta}
\label{e2}
.\end{equation}
\noindent
Substituting Eq.\,\ref{e2} into Eq.\,\ref{e1}, we obtain
\begin{equation}
F(i,\theta,\gamma) = \frac{3\,\cos^2{\gamma} - 1}{2} + \frac{\cos{\theta}\,\sin{2\,\gamma}}{2\,\tan{i}}
\label{e3}
,\end{equation}
from which we deduce
\begin{equation}
F_{\rm max} (i,\gamma) = \frac{3\,\cos^2{\gamma} - 1}{2} + \frac{\sin{2\,\gamma}}{2\,\tan{i}}
\end{equation}
and 
\begin{equation}
F_{\rm min} (i,\gamma) = \frac{3\,\cos^2{\gamma} - 1}{2} - \frac{\sin{2\,\gamma}}{2\,\tan{i}}
\end{equation}
as the maximum and minimum (relative) contributions for the rotational terms to the apsidal motion rate. 

\begin{figure}
\begin{center}
\includegraphics*[width=0.45\textwidth,angle=0]{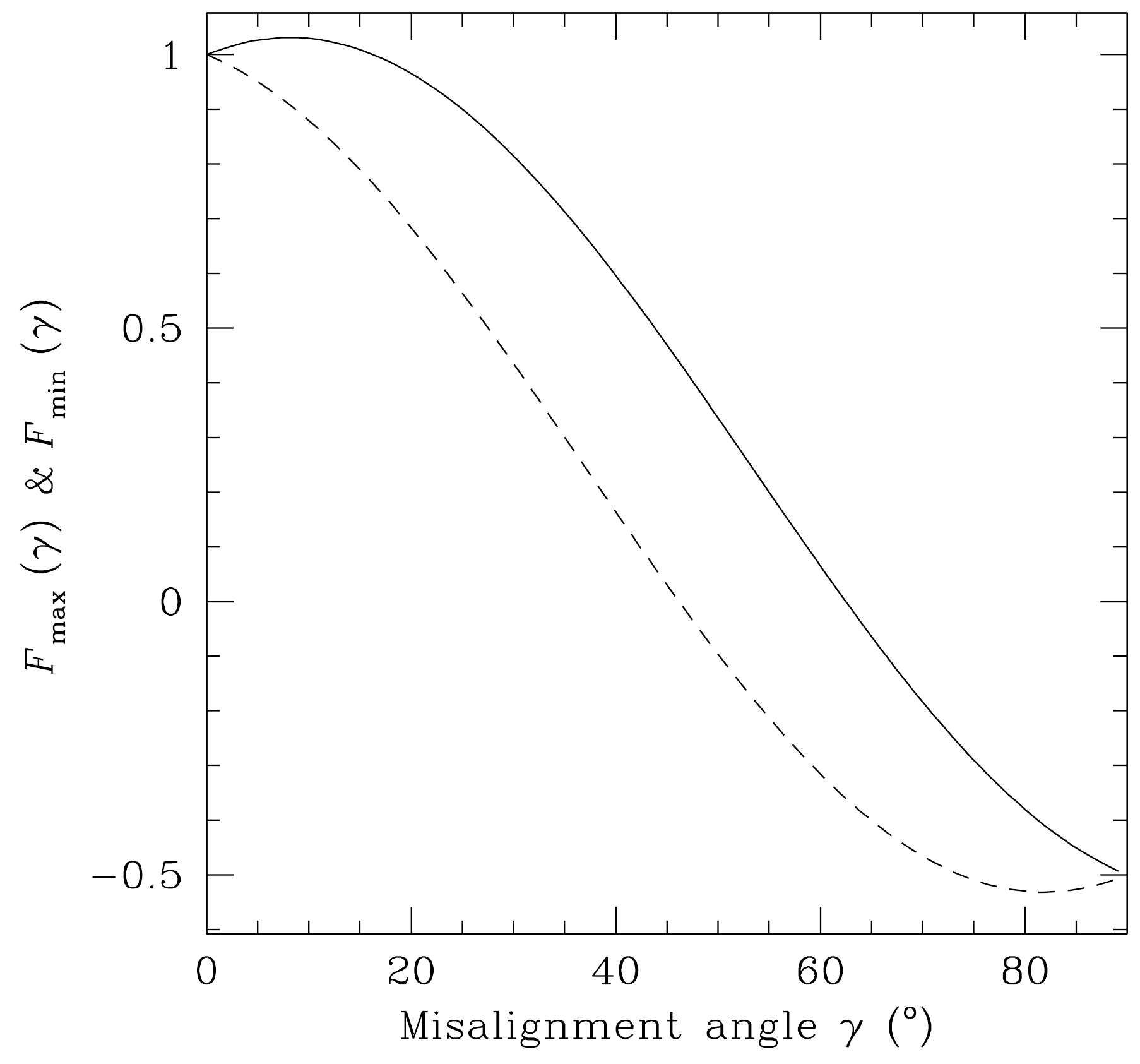}
\end{center}
\caption{Maximum (solid curve) and minimum (dashed curve) values of the misalignment correction factor $F$ (see Eq.\,\ref{e1}) as a function of misalignment angle $\gamma$. \label{misalign}}
\end{figure}

In Fig.\,\ref{misalign} we show these curves for the specific case of HD~152218 (with $i$ = 66.3$^{\circ}$); the area between the two curves shows the possible range of values of $F$ (as in Eq.\,\ref{e1}) as a function of the misalignment angle. The highest possible rotational contribution is reached at a non-zero misalignment angle $\gamma _{max}$, which can be trivially computed
from
\begin{equation}
\tan{2\,\gamma_{\rm max}} = \frac{2}{3\,\tan{i}}
.\end{equation}
However, this maximum contribution exceeds the value under perfect alignment only by 3\% and for most parameters the rotational contribution to the apsidal motion rate is diminished compared to the situation of perfect alignment. Hence, if we were to assume a misalignment between the rotational and orbital axes, we would reduce the predicted $\dot{\omega}$ as a function of age. As a result, such a misalignment would increase the age estimate based on the theoretical rate of apsidal motion. 
\begin{figure*}
\begin{center}
\includegraphics*[width=0.45\textwidth,angle=0]{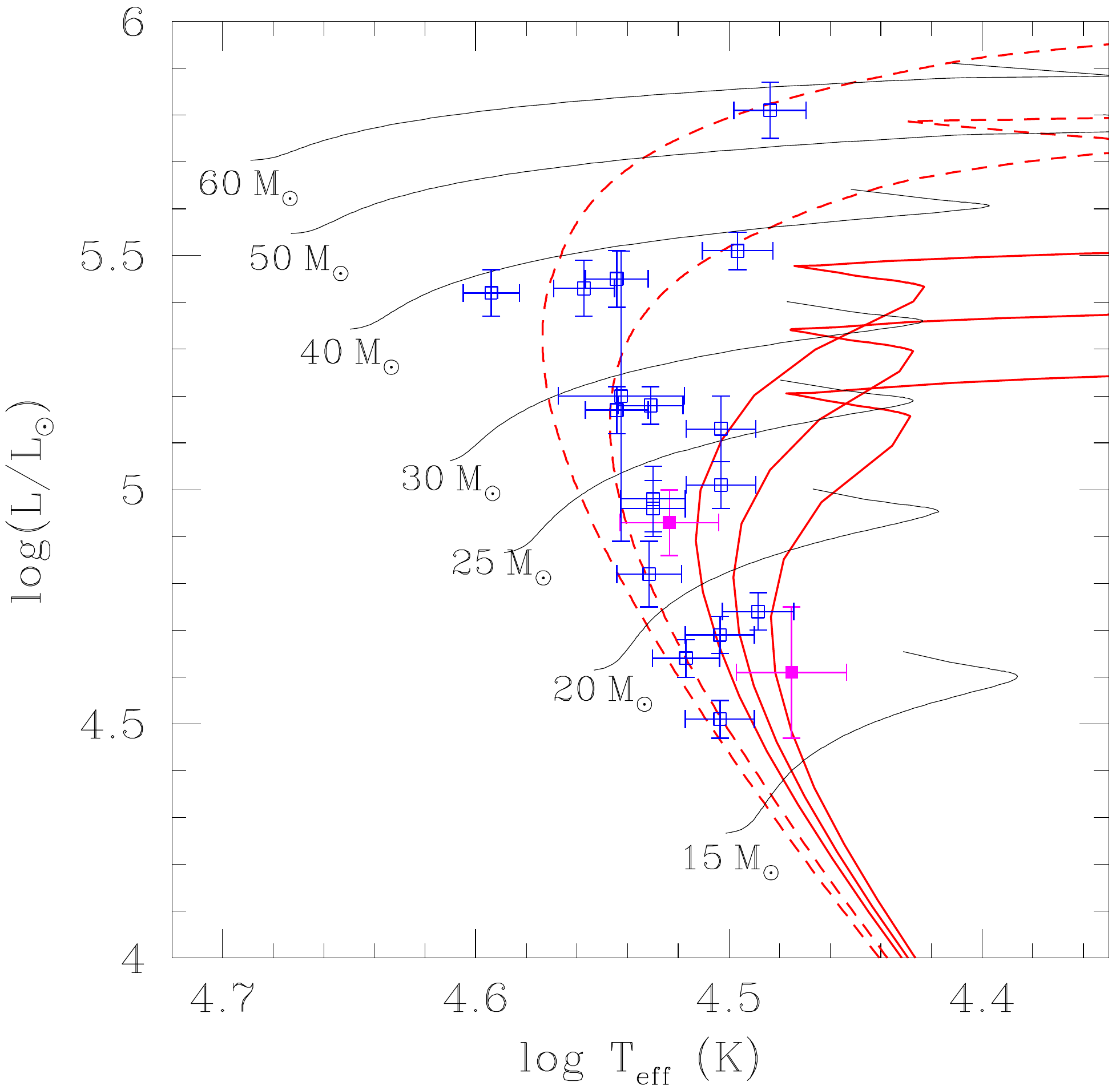}
\includegraphics*[width=0.45\textwidth,angle=0]{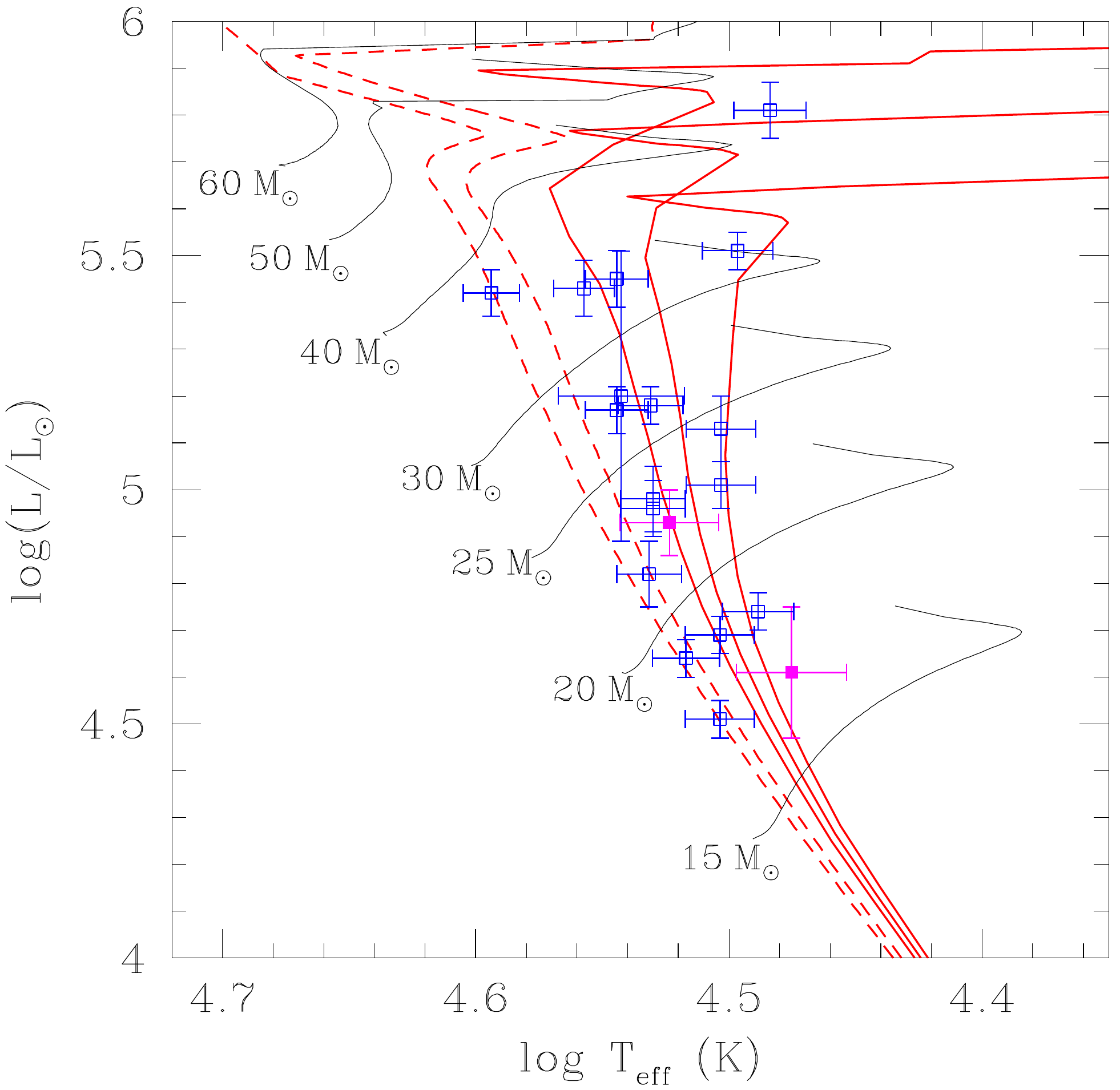}
\end{center}
\caption{Position of the O-type stars of NGC~6231 in a Hertzsprung-Russell diagram, accounting for the binarity. The components of HD~152218 are shown by the filled magenta symbols. Evolutionary tracks (identified with their initial mass) for $Z=0.014$ are taken from \citet{Ekstroem}. The red lines yield isochrones for $\log{\rm age} = 6.5$ and $6.6$ (dashed lines, i.e.\ ages of 3.2 and 4.0\,Myr, from left to right), and $\log{\rm age} =  6.71, 6.76$ and $6.81$ (continuous lines, i.e.\ ages of 5.1, 5.8 and 6.5\,Myr, from left to right). The latter isochrones bracket the range of ages inferred from the apsidal motion of HD~152218. The leftmost panel corresponds to evolutionary models without rotation, whilst the rightmost panel yields the tracks and isochrones for stars rotating at 40\% of their critical velocity. \label{HRD}}
\end{figure*}

To conclude this discussion about a possible misalignment of the rotational axes, we briefly consider the precession of the rotation axes that would result for such a situation \citep{Kopal1,Kopal2,Kopal3,Alexander}. From an observational point of view, precession would manifest itself through slow variations of the $v\,\sin{i_{\rm rot}}$ values. Using the formalism of \citet[][his Eq.\,5.46]{Kopal3} along with our theoretical values of $k_2$ and the moments of inertia derived from the CLES models, we estimate precession periods in the range 14 -- 19\,yr for the range of possible ages inferred in Sect.\,\ref{theory}. The fact that the {\it IUE} and FEROS spectra, taken about eight years apart, indicate consistent values of $v\,\sin{i_{\rm rot}}$ and the lack of conspicuous variations of the line widths over the five years covered by the FEROS data suggest that any possible misalignment of the rotation axes probably is rather modest. 

We therefore cannot identify any obvious bias in our result. Support for an age of about 6\,Myr comes from the photometric light curve. Although the agreement between the stellar radii inferred from the photometric light curve and the theoretical radii is certainly not perfect, the stellar radii inferred from the solution of the light curve clearly favour an age significantly in excess of 3\,Myr. 

Finally, to check whether such an age is consistent with the general properties of the population of massive stars in NGC~6231, we reconsidered all the O-type cluster members (except for the companion of the Wolf-Rayet binary WR~79) listed by \citet{NGC6231}. For the known binary systems, we corrected the magnitudes for the contribution of the companion. Whenever a dedicated spectroscopic study was available, we adopted the brightness ratio established in this study. For all other cases, we estimated the brightness ratio based on the preliminary spectral types listed by \citet{NGC6231}. We also adopted a distance modulus of $11.07 \pm 0.04$ and $R_V = 3.3$ \citep{Sana}. Effective temperatures and bolometric corrections were taken from the calibrations of \citet{Martins} and \citet{MP}. Except for cases where the uncertainties on spectral types are large (weakly detected secondary components of long-period binaries), we assumed an uncertainty of half a spectral subtype both in temperature and bolometric correction. The location of the stars in a Hertzsprung-Russell diagram is shown in Fig.\,\ref{HRD} along with single-star evolutionary tracks of \citet{Ekstroem} for $Z=0.014$, both without rotation (left panel) and with a 40\% critical rotation. This comparison shows that the O-type stars seem to span a range of ages, roughly between 3 and 6.3\,Myr. Our values of the age of HD~152218 as determined in Sect.\,\ref{theory} are towards the higher end of this range, but are consistent with the range of ages inferred with the isochrones.

\section{Conclusion \label{conclusions}}
We have re-analysed the massive binary HD~152218 in the young open cluster NGC~6231. Spectral disentangling allowed us to reconstruct individual spectra and hence to derive stellar temperatures and gravities comparing the individual spectra with CMFGEN model atmospheres. Photometric data were used to constrain the orbital inclination and the Roche-lobe filling factors. Radial velocity data from the literature allowed us to establish a rate of apsidal motion of $(2.04^{+.23}_{-.24})^{\circ}$\,yr$^{-1}$ , corresponding to a period of 176 years. Comparison of this rate with the predictions of stellar structure models yields an age of $5.8 \pm 0.6$\,Myrs for HD~152218. This value is towards the higher end of the range of ages of the isochrones that encompass the location of the O-star population of NGC~6231 in a Hertzsprung-Russell diagram.



\section*{Acknowledgements}
The Li\`ege team acknowledges support through an ARC grant for Concerted Research Actions, financed by the Federation Wallonia-Brussels, from the Fonds de la Recherche Scientifique (FRS/FNRS), including an FRS/FNRS Research Project (T.0100.15), as well as through an XMM PRODEX contract (Belspo).

\end{document}